\documentclass[pre,twocolumn,amsmath,amssymb,floatfix,10pt]{revtex4} 
\usepackage{graphicx,import}
\usepackage{epstopdf}
\usepackage{gensymb}
\usepackage{color}
\usepackage{amsmath}
\usepackage{titlesec}
\usepackage{natbib}

\begin{document}

\title{Pitch Perfect: How Fruit Flies Control their Body Pitch Angle}

\author{Samuel C. Whitehead$^{1*}$, Tsevi Beatus$^{1*}$, Luca Canale$^2$, and Itai Cohen$^1$}
\affiliation{$^1$Department of Physics, Cornell University, Ithaca, New York 14853, USA; $^2$D\'{e}partement de M\'{e}canique, \'{E}cole Polytechnique 911128, Palaiseau, France. \\ $^*$ Equal contributors }

\date{\today}

\begin{abstract}
Flapping insect flight is a complex and beautiful phenomenon that relies on fast, active control mechanisms to counter aerodynamic instability. To directly investigate how freely-flying \emph{D. melanogaster} control their body pitch angle against such instability, we perturb them using impulsive mechanical torques and film their corrective maneuvers with high-speed video. Combining experimental observations and numerical simulation, we find that flies correct for pitch deflections of up to 40$\degree$ in 29 $\pm$ 8 ms by bilaterally modulating their wings' front-most stroke angle in a manner well-described by a linear proportional-integral (PI) controller. Flies initiate this corrective process after only 10 $\pm$ 2 ms, indicating that pitch stabilization involves a fast reflex response.  Remarkably, flies can also correct for very large-amplitude pitch perturbations--greater than 150\degree--providing a regime in which to probe the limits of the linear-response framework.
Together with previous studies regarding yaw and roll control, our results on pitch show that flies' stabilization of each of these body angles is consistent with PI control. 

\end{abstract}

\maketitle

\section{Introduction}
\label{sec:intro}

From walking humans to flying insects, many fascinating forms of bio-locomotion are contingent upon robust stabilization control. Implementing this control is particularly difficult in the case of small, flapping-wing insects, since flapping flight is inherently subject to rapidly-divergent aerodynamic instabilities \cite{TaylorThomasJEB2003, SunBumblebeeJEB2005, TaylorZibkowski2005, SunDynamicStability2007, LiuCFD2010, FaruqueHumbert2010a, ZhangSunLateral2010, ZhangSunLateralControl2011, GaoCFD2011, WoodVerticalFlight2011, RistrophPitchInterface2013, XuSunCFD2013, SunRevModPhys2014}. As such, flying insects have evolved stabilization techniques relying on reflexes that are among the fastest in the animal kingdom \cite{BeatusInterface2015} and robust to the complex environment that insects must navigate \cite{CombesPNAS2009, DickersonRaindrop2012, RaviJEB2013, OrtegaJEB2013, VanceGust2013}. 

In particular, pitching instability is a prominent obstacle for flight control in flapping insects. Analytical and numerical modeling suggest that, for many two-winged insects (e.g. flies), periodic flapping couples with longitudinal body motion to produce rapidly-growing oscillations of the body pitch angle \cite{SunDynamicStability2007,RistrophPitchInterface2013,ChangPNAS2014,SunRevModPhys2014}. This oscillatory instability can be understood as the result of differential drag on the wings due to longitudinal body motion \cite{SunDynamicStability2007, RistrophPitchInterface2013}. For example, if a fly pitches down while hovering, its re-directed lift propels its body forward, causing an increased drag on the wings during the forward stroke relative to the backward stroke. Because dipteran wings are attached to their bodies above the center of mass, this drag asymmetry generates a torque that pitches the fly up. Rather than acting as a restoring torque, the drag--together with the body inertia--pitches the fly up, beyond its initial pitch orientation. The fly then begins to move backwards, and oscillation ensues in the opposite direction. This mechanism results in an undulating instability of the body pitch angle, which doubles over a timescale of $\sim9$ wing-beats \cite{SunRevModPhys2014}. Mitigating the effects of this instability requires flies to actively adjust their wing motion on time scales faster than the growth of these oscillations. 

Our work builds upon an already rich corpus of literature on insect flight control, a sizable portion of which addresses the pitch degree of freedom. Experimental studies subjecting tethered insects to both mechanical pitching perturbations and visual pitching stimuli \cite{ZankerIIIControl1990, NalbachNonOrthogonal1994, DickinsonHaltere1999, ShermanDickinsonJEB2003, ShermanDickinsonJEB2004,TaylorThomasJEB2003} have elucidated stereotyped kinematic responses for pitch correction, including manipulation of wing stroke angle, stroke plane orientation, wing-beat frequency, and body configuration. However, tethered insects do not constitute a closed-loop feedback system, since changes to their wing kinematics do not affect their body orientations. Moreover, in the case of tethered flies, it has been shown that the wing kinematics are qualitatively different than those in free-flight \cite{FryJEB2005, BenderVisual2006}. Thus, free-flight studies are necessary for a comprehensive understanding of pitch control. Significant analysis has been performed on freely-flying insects executing voluntary maneuvers \cite{EnnosJEB1989, FryScience2003, RistrophJEB2009} or responding to visual stimuli \cite{ChengHedrickJEB2011, WindsorInterface2014, MuijresScience2014}, but the general challenge of systematically inducing mechanical perturbations on untethered insects has traditionally been a barrier to the study of stabilization reflexes. Some notable exceptions to this include methods of mechanical perturbation using air-flow vortices \cite{CombesPNAS2009, RaviJEB2013, OrtegaJEB2013} or gusts of wind \cite{VanceGust2013}. However, such fluid-impulse methods are difficult to tune, and are thus not ideal for of inducing the fast, precise mechanical perturbations that are required for a quantitative understanding of pitch control. 

\begin{figure}[h]
	\centering
		\hspace*{-1cm}\includegraphics[scale=.8]{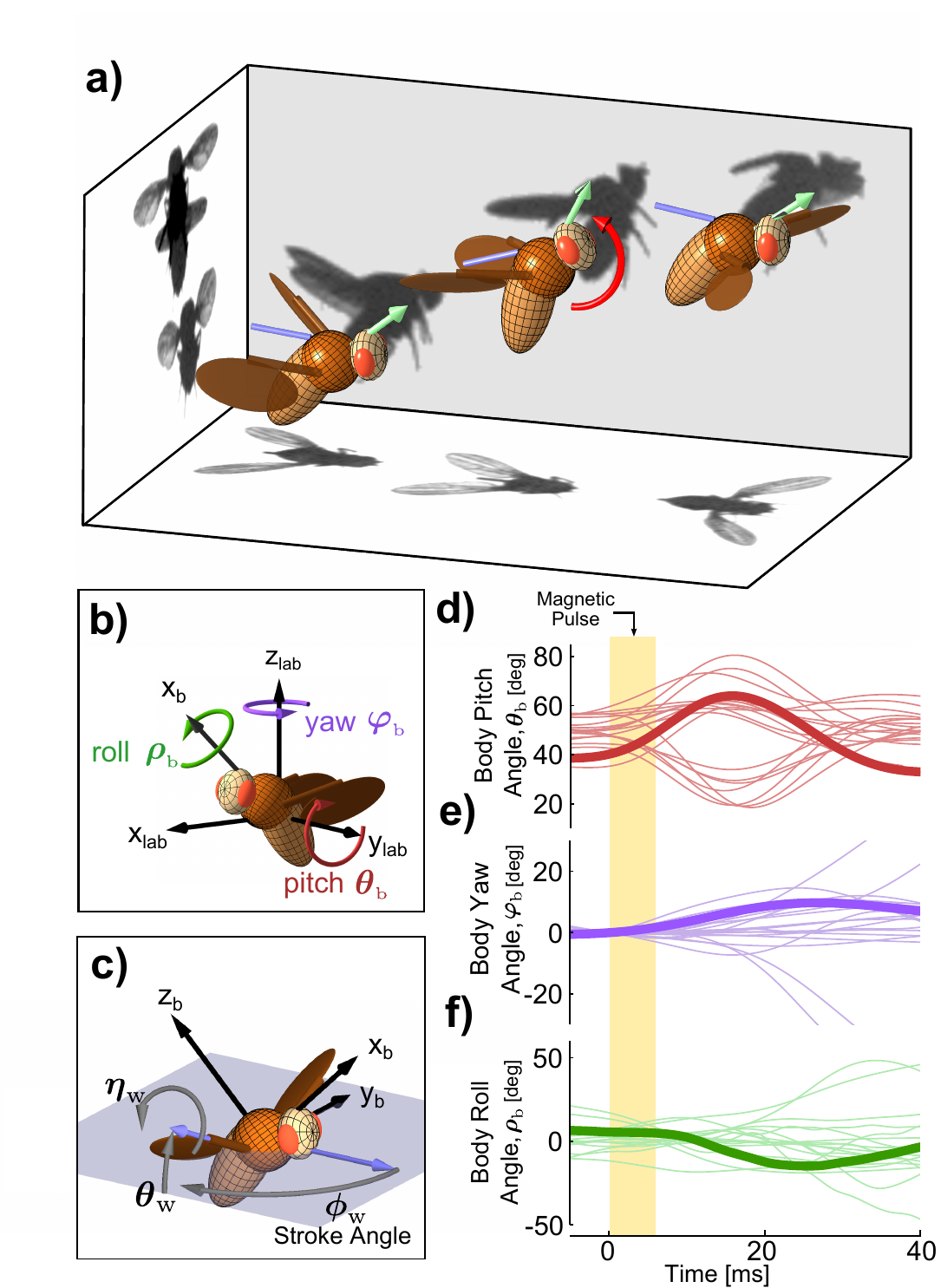}
		\caption{Pitch perturbation and correction. (a) snapshots and a 3D model reconstruction from a representative pitch up event at $t =$ -25.5, 13, 56.5 ms. For the full video of this event, see Movie S1. The middle snapshot ($t$ = 13 ms) corresponds to the maximum pitch up deflection. Ferromagnetic pin is false-colored blue, the fly's long body axis is given by the green arrows, and the red arrow indicates the direction of the perturbation. (b) and (c) definitions for the body and wing Euler angle coordinates, respectively. $\varphi_\text{b}$, $\theta_\text{b}$, and $\rho_\text{b}$ indicate the body Euler angles, while $\phi_\text{w}$, $\theta_\text{w}$, and $\eta_\text{w}$ indicate wing Euler angles; the stroke plane is shown in gray (c). Also shown are the lab (b) and body (c) frames of reference. (d-f) time series of body Euler angles for 18 perturbation events, with the highlighted curves corresponding to the event from (a). The yellow bar in (d-f) gives the timing of the magnetic pulse (0-5.8 ms). Body angles in (d-f) are spline-smoothed from raw data, with body yaw in (e) shifted by its value at $t = 0$.}
		\label{fig:bodykinematics}
\end{figure}

To achieve the necessary speed and precision for measurements of body pitch control, we use a perturbation scheme that has previously been applied to analyzing control of the yaw \cite{RistrophPNAS2010} and roll \cite{BeatusInterface2015} degrees of freedom. We mechanically perturb free-flying \emph{D. melanogaster} by gluing small magnetic pins to their dorsal thoracic surfaces and applying short bursts (5-8 ms) of a vertical magnetic field that pitches their bodies up or down. As the flies correct their orientation, we measure their body and wing kinematics using high-speed video (Figure \ref{fig:bodykinematics}a). 

We recorded perturbation events with amplitudes typically ranging 5-40$\degree$ for both pitching up and pitching down. For these perturbations, flies recover 90\% their pitch orientation within $\sim$30 ms. Moreover, we find that the corrective process is initiated $\sim$2 wing-beats ($\sim10$ ms) after the onset of the impulsive torque; such a short latency time indicates that this corrective process is a reflexive behavior largely governed by input from the halteres, the flies' rate-gyro-like mechanical sensing organs \cite{PringleGyroscopic1948, NalbachNonOrthogonal1994, DickinsonHaltere1999}. 

To generate corrective pitching torques, flies bilaterally modulate their wings' front-most stroke angle, i.e. they flap their wings more or less in the front to pitch up or down, respectively. This corrective mechanism is in general agreement with previous findings on active body pitch stabilization in fruit flies \cite{ZankerIIIControl1990, DickinsonHaltere1999, TaylorMechanics2001, ChangPNAS2014}. We show that flies' modulation of front stroke angle is well-modeled by a linear, continuous, proportional-integral (PI) controller with a time delay, $\Delta T = 6 \pm 1.7$ ms (mean $\pm$ standard deviation). Our results indicate that pitch control in fruit flies is an extremely fast and robust process, which can be accurately modeled by a simple controller for a wide range of perturbation amplitudes. Moreover, we find that flies are capable of correcting for pitch deflections of 150$\degree$ or more, a perturbation regime in which the linear controller theory begins to break down. Together with previous results on how flies control yaw \cite{RistrophPNAS2010} and roll \cite{BeatusInterface2015}, the analysis of pitch control presented here addresses a missing piece in our understanding of how flies control each of their body angles individually.

\section{Methods}
\label{sec:methods}

\subsection{Fly Preparation}
\label{subsec:flyprep}
\begin{figure*}[t]
	\centering
		\includegraphics[scale = .9]{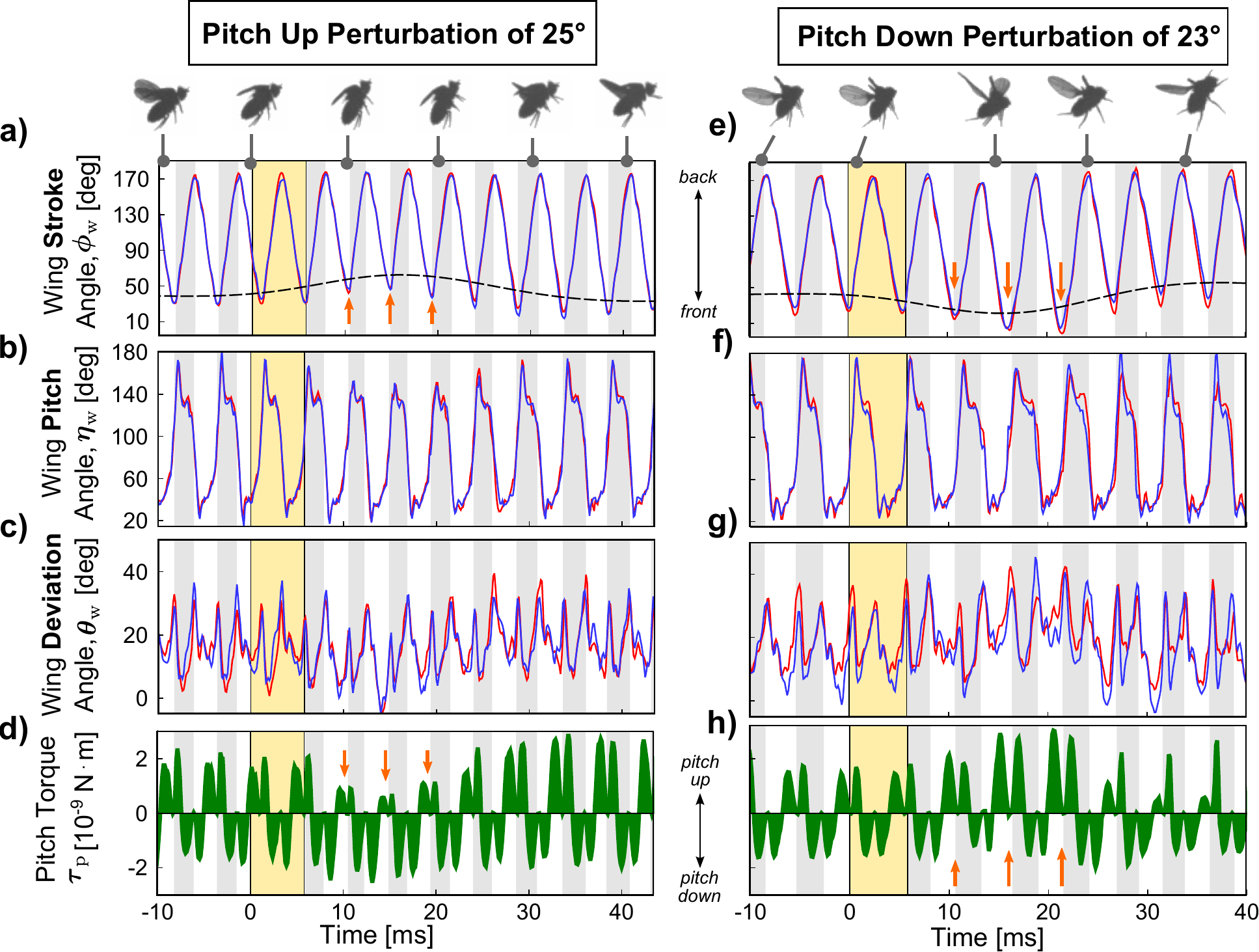}
		\caption{Wing kinematics and aerodynamic torques for two representative perturbation events. (a-d) correspond to a pitch up perturbation with maximum amplitude 25$\degree$; (e-h) correspond to a pitch down perturbation with maximum amplitude 23$\degree$. Plotted wing kinematics as a function of time include: stroke angle (a),(e); wing pitch angle (b),(f); and deviation angle (c),(g). Instantaneous aerodynamic torques about the flies' pitch axis are given in (d),(f). Orange arrows in (a) and (e) highlight corrective front strokes, with corresponding arrows in (d) and (h) highlighting the changes in pitch torque resulting from the corrective kinematics. Images above (a) and (d) show side views of the flies (raw data) at different points during the movie, illustrating the changes in body pitch that accompany a perturbation. White and gray bars indicate forward and back strokes respectively; yellow bar corresponds to the perturbation duration (0 - 5.8 ms).}
		\label{fig:wingkinematics}
\end{figure*}

We perform each experiment using common fruit flies (\emph{Drosophila melanogaster}, females) from an out-bred laboratory stock. Individual flies are anesthetized at 0-4\degree C, at which point we carefully glue  1.5-2 mm long, 0.15 mm diameter ferromagnetic pins to their notum (dorsal thoracic surface), oriented so that the pin lies in the flies' sagittal plane. The pin is shown in Figure \ref{fig:bodykinematics}a (false-colored blue) and Figure \ref{fig:wingkinematics} (images). Control experiments with untreated flies show that the addition of the pin does not qualitatively alter flies' flight kinematics. When attached, the pins add $\sim$20\% to the mass and pitch moment of inertia of the fly, which falls within the range of their natural body mass variations. Moreover, the pin contributes negligibly to off-diagonal components of the flies' inertia tensors, and therefore does not introduce any coupling between the rotational degrees of freedom of the body. The primary effect of the added pin mass is a dorsal shift of $\sim$0.2 mm in the flies' center of mass, which must be accounted for during calculations of aerodynamic torque.

\subsection{Videography and Mechanical Perturbation}
\label{subsec:pert}
Once 15-30 flies have been prepared as above, we release them into a transparent cubic filming chamber of side length 13 cm. On the top and bottom of the chamber are attached two horizontally-oriented Helmholtz coils, which produce a vertical magnetic field. The central region of the chamber is filmed by three orthogonal high-speed cameras (Phantom V7.1) at 8000 frames per second. The cameras are calibrated using a direct linear transformation scheme detailed in \cite{LourakisSBA2009, TheriaultJEB2014}. When flies enter the filming volume, an optical trigger simultaneously signals the cameras to record and supplies a 5-8 ms current pulse to the Helmholtz coils. We varied both the strength and duration of the magnetic pulse produced by the coils across experiments. Maximal field strengths reached $\sim$10$^{-2}$ Tesla, and most experiments were performed with a pulse that lasted 5.8 ms. The magnetic pulse exerts a torque on the ferromagnetic pin, pitching the fly either up or down.

Of the movies collected using the above method, we selected 18 to analyze in full; 16 additional movies were partially analyzed to collect more data on pitch correction time and to observe correction for very large-amplitude perturbations ($\sim$ 150\degree). We chose movies to fully analyze based on the criteria that i) the time window during which the fly is in the field of view for all three cameras is sufficiently long to observe pre- and post-perturbation kinematics, ii) the perturbation primarily affects the fly's pitch orientation, iii) the fly is not performing any maneuver other than correction, and iv) we sample a wide range of perturbation magnitudes for both pitching up and pitching down across our data set. To glean kinematic data from the raw footage, we use a custom-developed image analysis algorithm detailed in \cite{RistrophJEB2009, BeatusInterface2015}. This 3D hull reconstruction algorithm provides a kinematic description of 12 degrees of freedom for the fly (body orientation and center of mass position, as well as three Euler angles for each wing). 

\section{Results}
\label{sec:results}

\subsection{Body and Wing Kinematics During Pitch Correction} 
\label{subsec:kinematics}
Representative kinematics for perturbation events are shown in Figures \ref{fig:bodykinematics} and \ref{fig:wingkinematics}. Figure \ref{fig:bodykinematics}b shows definitions for the body Euler angle coordinates--yaw ($\varphi_\text{b}$), roll ($\rho_\text{b}$), and pitch ($\theta_\text{b}$). Figures \ref{fig:bodykinematics}d-f show time series of these Euler angles before and after the application of a 5.8 ms magnetic pulse (yellow strip) for 18 perturbation events. Before the perturbation, flies typically maintain a pitch angle of roughly 50$\degree$. Perturbations deflect the pitch angle by as much as 40$\degree$ either up or down. While the flies' yaw and roll angles are sometimes altered by the perturbation, pitch is the most consistently and significantly affected degree of freedom immediately following the application of the pulse (at $t\approx6$ ms).  Highlighted curves in Figures \ref{fig:bodykinematics}d-f show an event in which the fly was pitched up by 25$\degree$, attaining its maximal angular deflection at 15 ms after the onset of the perturbation. By 29 ms it has corrected for 90$\%$ of the pitch deflection. The maximum pitch velocity due to the perturbation was 2400 \degree/s. For the full movie of this perturbation event, see Movie S1.

The wing kinematics for two representative perturbations, one in which the fly is pitched up by 25$\degree$ (the same event highlighted in Figures \ref{fig:bodykinematics}d-f) and another in which the fly is pitched down by 23$\degree$, are shown in Figure \ref{fig:wingkinematics}. Wing Euler angles--stroke ($\phi_\text{w}$), pitch ($\eta_\text{w}$), and deviation ($\theta_\text{w}$)--are defined in Figure \ref{fig:bodykinematics}c. In general, the wing kinematics we observe following the perturbation are left/right symmetric. Hence, these kinematics can be attributed to pitch correction, since both yaw \cite{RistrophPNAS2010} and roll \cite{BeatusInterface2015} correction require left/right asymmetric wing motion. For the pitch up event, $\sim$10 ms, or 2 wing-beats, after the onset of the perturbation, the minima of the wing stroke angles shift upward for both the left and right wing, as indicated by the orange arrows in Figure \ref{fig:wingkinematics}a. During a given wing-beat cycle, the minimum of the stroke angle for each wing corresponds to its front-most position. Since pitch correction is left/right symmetric, we refer to the average of the front-most positions for the left and right wings as the front stroke angle, $\phi_\text{w}^{\text{front}}$. By 15 ms, the fly in Figures \ref{fig:wingkinematics}a-d has increased its $\phi_\text{w}^{\text{front}}$ from its pre-perturbation value by $\sim$25\degree. Physically, this means that the fly is significantly reducing the amplitude of its ventral stroke, i.e. flapping \emph{less forward}. The duration of this increase in $\phi_\text{w}^{\text{front}}$ is 3 wing-beats. We do not observe any shifts in the fly's back-most stroke angle, $\phi_\text{w}^{\text{back}}$, during the correction maneuvers.

Conversely, for the pitch down event in Figures \ref{fig:wingkinematics}e-h, $\sim$10 ms after the perturbation onset the pitched-down fly begins to \emph{decrease} its $\phi_\text{w}^{\text{front}}$. This corresponds to the fly increasing the amplitude of its ventral stroke, i.e. flapping \emph{further forward}. Again, there appears to be little to no change in the back stroke angle. Put together, these results are consistent with previous kinematic measurements \cite{ZankerIIIControl1990, TaylorMechanics2001, DickinsonHaltere1999, RistrophPitchInterface2013} and suggest that flies modulate their front stroke angle to produce corrective pitch torques, increasing $\phi_\text{w}^{\text{front}}$ to pitch themselves down, and decreasing it to pitch themselves up. 

\subsection{Aerodynamic Forces and Torques}
\label{subsec:aero}
\begin{figure}
		\includegraphics[scale = .70]{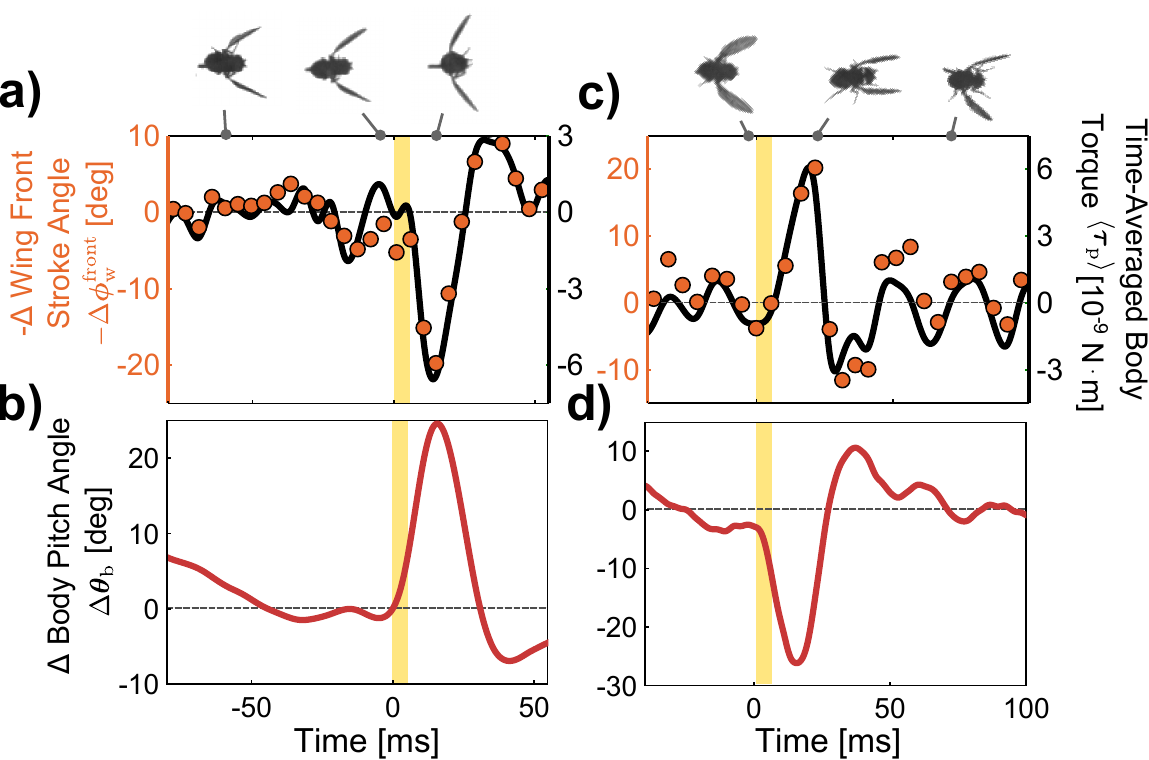}
		\caption{Mechanism for generating corrective pitch torques. Data in (a),(b) corresponds to the pitch up event shown in Figures \ref{fig:bodykinematics}d-f and \ref{fig:wingkinematics}a-d, while (c),(d) correspond to the pitch down event shown in Figures \ref{fig:wingkinematics}e-h. (a),(c) wing-stroke-averaged pitch torque (black line, spline interpolated) and negative front stroke angle deviation (orange dots) vs time. Front stroke angle deviation is defined by $\Delta\phi_\text{w}^\text{front}(t) = \phi_\text{w}^\text{front}(t) - \phi_\text{w}^\text{front}(0)$. We flip the sign of $\Delta\phi_\text{w}^\text{front}$ to illustrate correlation with torque. (b),(d)  body pitch angle (red line) vs time. As in previous plots, the yellow strip in (a)-(d) corresponds to the duration of the magnetic pulse. }
		\label{fig:torquecomp}
\end{figure}

Intuitively, the relationship between front stroke angle and pitching torque can be understood as follows. To within a good approximation, the net aerodynamic force generated by a flapping wing is directed perpendicular to the wing's surface \cite{DickinsonScience1999, SaneJEB2001}, so that portions of the wing stroke during which the wing is in the front half of the stroke plane ($\phi_\text{w} \le$ 90\degree) generate pitch up torques, while portions in the back half ($\phi_\text{w} \ge$ 90\degree) generate pitch down torques. During non-maneuvering flight, these torques cancel over a wing stroke. By biasing a wing stroke so that it spends a smaller fraction of the stroke period in the forward position, a smaller pitch up torque is generated during that cycle, such that the net pitch torque will be directed downward. Conversely, by increasing the front stroke angle, and thus increasing the portion of the stroke spent in the front position, flies can generate a net pitch-up torque over the course of a full stroke. This can be observed in Figures \ref{fig:wingkinematics} (orange arrows) and \ref{fig:torquecomp}, in which \emph{active} adjustments to front stroke angle result in net corrective pitching torques over the course of individual wing-beats. Modulating the balance of two opposing forces or torques is used by flies in other maneuvers such as forward \cite{RistrophPaddling2011} and sideways \cite{RistrophJEB2009} flight, as well as yaw \cite{RistrophPNAS2010} and roll \cite{BeatusInterface2015} maneuvers, and appears in other bio-locomoting systems, such as knifefish \cite{sefati2013mutually}.

To quantify the effect of changing the front stroke angle, we calculate the pitching torque generated by the wings during the maneuvers shown in Figure \ref{fig:wingkinematics} using the full 3D fly kinematics. To calculate the aerodynamic force generated by the wings, we used a quasi-steady aerodynamic force model that was previously calibrated on a mechanical, scaled-up fly model \cite{DickinsonScience1999, SaneJEB2001}. This model gives the lift ($F_\text{L}$) and drag ($F_\text{D}$) forces generated by the wings as:
\begin{align}
\centering
F_\text{L} &= \tfrac{1}{2}\rho_{0} S U_\text{t}^2 r^2_2 C_\text{L}(\alpha)  \label{eq:lift}\\
F_\text{D} &= \tfrac{1}{2}\rho_{0} S U_\text{t}^2 r^2_2 C_\text{D}(\alpha)  \label{eq:drag}
\end{align}
$C_\text{L}$ and $C_\text{D}$ are the wing's lift and drag coefficients respectively, and are given as functions of angle of attack ($\alpha$) by \cite{SaneJEB2001}; $S$ is the wing area, $r_2^2$ the non-dimensional second moment of wing area (given as 0.313 by \cite{ChengIEEE2009}), $\rho_{0}$ the density of air, and $U_\text{t}$ the wingtip velocity. Drag is directed anti-parallel to the wing tip velocity, and lift is perpendicular to both drag and the wing span vector. The total aerodynamic force is the vector sum of the lift and drag forces. While this is a simple method for calculating forces on flapping wings, we find that it quantitatively captures the relevant force production for both pre- and post-perturbation wing kinematics. We tested the effect of adding rotational forces to the aerodynamic model \cite{SaneJEB2001}, which should give the next largest contribution to force, but found negligible changes to our calculation results.

From these lift and drag forces, we calculate aerodynamic torques exerted by the wings on the body, shown in Figures \ref{fig:wingkinematics}d,h. The moment arm for the torque is given by the vector from the fly's center of mass to the wing's center of pressure, assumed to be in the chord center, 70\% along the length of the wing's span. In Figure \ref{fig:wingkinematics}d, during the active correction, the fly's wings generate a net downward pitching torque to oppose the perturbation. The main effect we observe is that, during a given wing cycle the wings generate less upward torque, resulting in a net downward bias for the torque on the body. This net pitch down torque is highlighted by the orange arrows in Figure \ref{fig:wingkinematics}d. Similarly, the pitched down fly in Figure \ref{fig:wingkinematics}h generates net upward pitching torque during active correction, also highlighted by orange arrows. The corrective torques for these two events are well-correlated with the measured modulations of front stroke angle (Figure \ref{fig:torquecomp}a,c), a trend that we observe across all perturbation events. 

Interestingly, after the flies generate a corrective torque for 2-3 wing-beats, we also observe a few wing-beats in which they generate net torque in the opposite direction (Figure \ref{fig:wingkinematics}d,h and \ref{fig:torquecomp}a,c). As with the initial corrective torque, this subsequent counter-torque arises from modulations of the front stroke angle, evident in Figures \ref{fig:torquecomp}a,c. The counter-torque acts to brake the corrective pitching motion, mitigating the overshoot in body pitch angle caused by the initial correction response. This allows for faster correction times, since the initial corrective maneuver can generate larger torques, and thus more quickly return the fly to pitch angles near its original orientation. In movies that allowed us to track the fly for long times after the perturbation, we observe that the front stroke angle and the net aerodynamic torque often oscillate with decaying amplitude and a period of $\sim$3-4 wing-beats. 

Importantly, passive damping of pitch motion contributes negligibly to the correction maneuvers we observe, since the characteristic passive decay time for pitch velocity is $\gtrsim 150$ ms, much longer than the entire correction maneuver \cite{Cheng2010Near}. Taken together, our results indicate that pitch correction for flies is an \emph{active} process involving modulation of $phi_\text{w}^\text{front}$.

\subsection{The Importance of Stroke Angle Relative to Other Degrees of Freedom}
\label{subsec:stroke}

\begin{figure}
		\includegraphics[scale = .8]{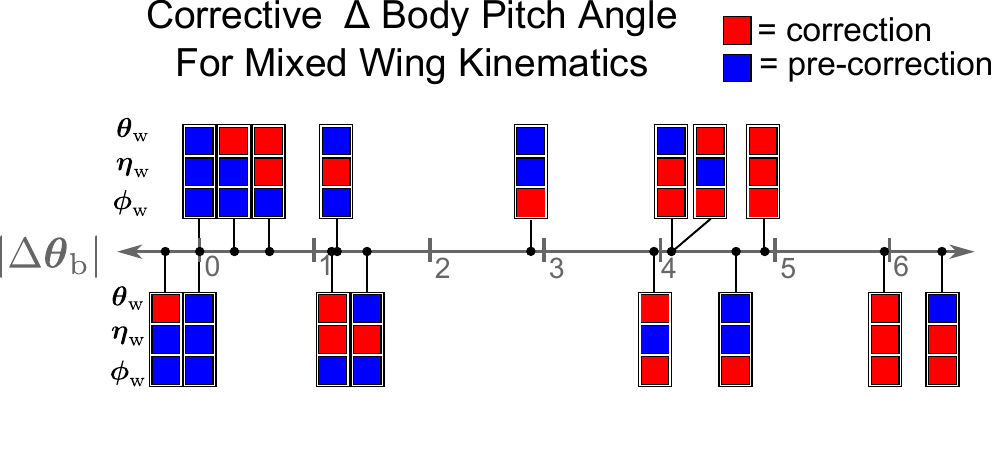}
		\caption{The magnitude of body pitch angle change resulting from different combinations of wing Euler angles for both a pitch up and pitch down event. Individual points correspond to unique combinations of pre- and post-perturbation wing angle kinematics, which are used with our quasi-steady aerodynamic model to calculate pitching torques, which in turn are used to estimate pitch angle deflection over a wing-stroke. Wing kinematics $\phi_\text{w}$ (stroke), $\theta_\text{w}$ (deviation), and $\eta_\text{w}$ (wing pitch) are defined in Figure \ref{fig:bodykinematics}c. Points above the axis correspond to kinematics taken from the pitch up event in Figures \ref{fig:wingkinematics} and \ref{fig:torquecomp}; points below the axis correspond to kinematics taken from the pitch down event in Figures \ref{fig:wingkinematics} and \ref{fig:torquecomp}.}
		\label{fig:leifLike}
\end{figure}

To assess the effect of front stroke angle modulation on body pitch correction, we calculate the changes to body pitch angle generated by changes in wing kinematics. We isolate the corrective effect of each wing kinematic variable by first identifying wing-strokes that correspond to both non-maneuvering (no net torque) and corrective flight. We then separate out the kinematic variables for each type of flight, and calculate the changes to body pitch angle resulting from different combinations of corrective and non-corrective kinematics, shown by the color codes in Figure \ref{fig:leifLike}. For example, a point with color combination blue-blue-red corresponds to a wing-beat with wing pitch and deviation angles taken from the corrective maneuver and the stroke angle taken from non-maneuvering flight. Our calculation uses the aerodynamic model detailed above to determine the torques produced by a given set of wing kinematics, assuming rigid wings attached to a stationary body at a point above the body center of mass. From the calculated torques, we determine the net change in body pitch angle over the course of each wing-stroke using numerical integration.  

We perform this analysis for all 8 possible combinations of wing kinematics for data from two different perturbation events, the pitch up and pitch down events in Figures \ref{fig:wingkinematics} and \ref{fig:torquecomp}. Points corresponding to combinations of wing kinematics from the pitch up and pitch down events are shown above and below the axis, respectively. In both cases, the sign for $\Delta\theta_\text{b}$ is chosen so that positive indicates a corrective rotation. The grouping of points in Figure \ref{fig:leifLike} indicates that body pitch correction is most closely associated with changes to wing stroke angle. The blue-blue-red point, corresponding to corrective stroke angle but non-maneuvering wing pitch and rotation, achieves at least 60$\%$ of the correction to body pitch angle, consistent with \cite{MuijresScience2014}. Moreover, the only combinations that give more than 30$\%$ of the total correction have corrective stroke angle (points of the form x-x-red). Changes to wing pitch or deviation can contribute $\sim 40\%$ to the corrective process, but cannot alone account for corrective body kinematics. These results motivate a minimal model for body pitch stabilization that considers only variations in front stroke angle to drive pitch correction.

\subsection{The Corrective Effect of Stroke Angle Over a Range of Perturbations}
\label{subsec:stats}
To further flesh out the relationship between front stroke angle and corrective torque, we plot the maximum measured corrective pitch acceleration generated by the fly in each maneuver as a function of the corresponding change in front stroke angle measured at that time, $\Delta\phi_\text{w}^\text{front}$ (Figure \ref{fig:stats}a). The maximum pitch acceleration was measured at the extremum of $\theta_\text{b}$, using a quadratic polynomial fit. The plot demonstrates a strong correlation between changes in front stroke angle and corrective acceleration (linear R$^2$ = 0.87). Consistent with the two maneuvers in previous figures, Figure \ref{fig:stats}a shows that, across our data set, flies increase $\Delta\phi_\text{w}^\text{front}$ (flap less forward) to pitch themselves down, and decrease $\Delta\phi_\text{w}^\text{front}$ (flap further forward) to pitch themselves up. 

The correlation between $\Delta\phi_\text{w}^\text{front}$ and corrective pitch acceleration in Figure \ref{fig:stats}a is also predicted by a calculation based on the quasi-steady aerodynamic force model in Equations \ref{eq:lift},\ref{eq:drag} (Figure \ref{fig:stats}a, gray line). To calculate the aerodynamic pitch torques, we use a simplified wing kinematic model similar to that in \cite{ChangPNAS2014} in which only the front stroke angle is varied (see Supplementary). We average the computed torques over a wing-beat, and divide by the moment of inertia to obtain pitch acceleration. The calculation relies only on the wing kinematics and fly morphology \cite{ChengIEEE2009}, and has no fitted parameters. Moreover, we verified that the resultant acceleration is insensitive to body pitch rotations, suggesting that passive damping contributes negligibly to correction \cite{Cheng2010Near}. The results of our calculation, shown in Figure \ref{fig:stats}a, quantitatively reproduce the measured pitch acceleration, further corroborating a model for pitch control that includes only modulation of $\phi_\text{w}^\text{front}$. 

To rule out an alternative corrective mechanism, based on modulation of back stroke angle, we plot corrective pitch acceleration as a function of $\phi_\text{w}^\text{back}$, shown in Figure \ref{fig:stats}b. We find no discernible correlation between these two variables. Calculating aerodynamic forces predicts that changes to the back stroke angle could produce corrective pitching torques in the same way that changes to front stroke angle do; the fact that we do not observe this in the data hints that morphological constraints favor modulation of the front stroke angle.  

\subsection{Correction Timescales}
\label{subsec:corrtimes}
\begin{figure}
	\centering
		\includegraphics[scale = .75]{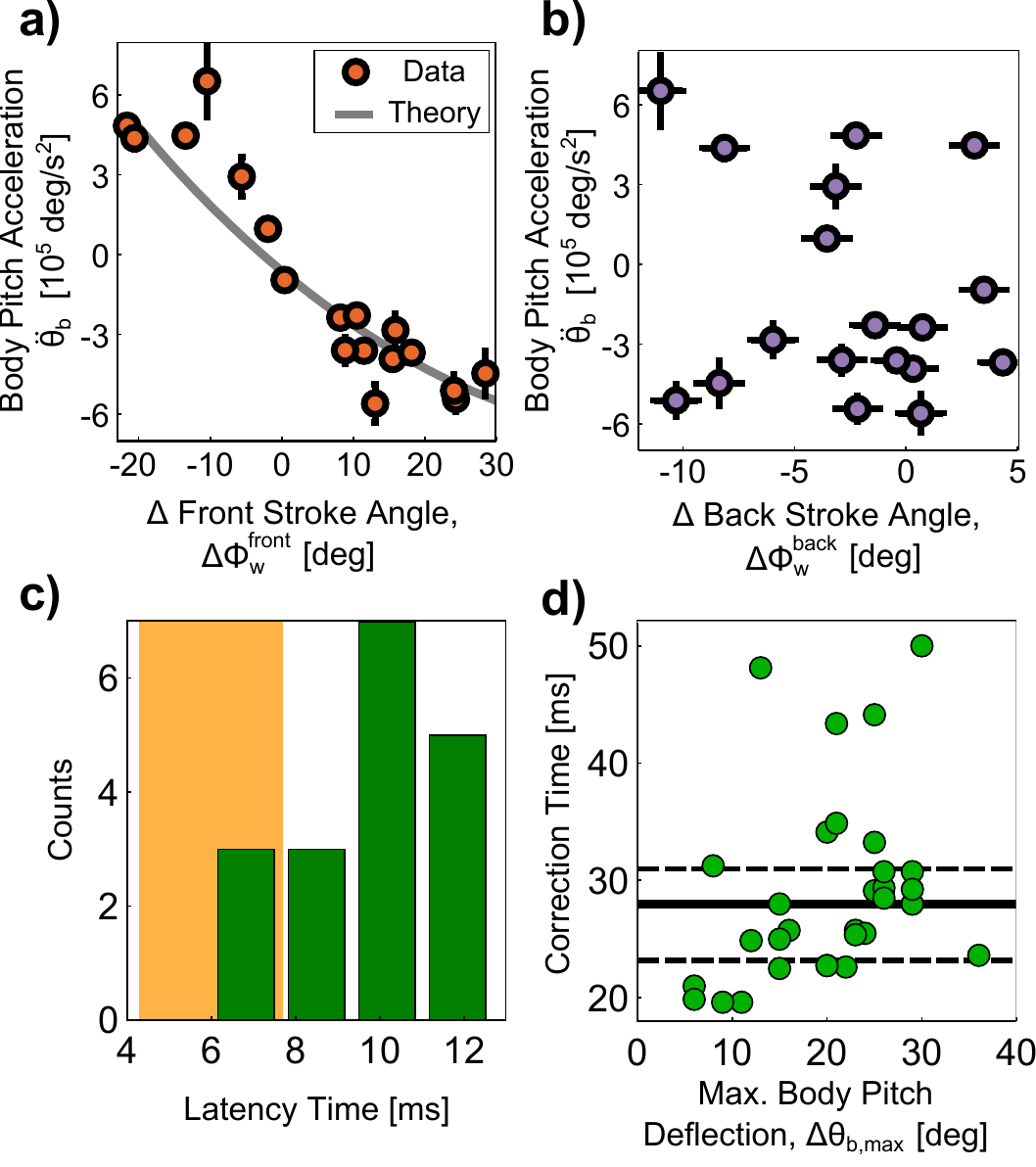}
		\caption{Statistics from many perturbation events. (a-b) Maximum corrective pitch acceleration generated by the fly as a function of change in front stroke angle and back stroke angle. The gray line in (a) is the calculated pitch acceleration for simplified wing kinematics. Note that the gray line has no fit parameters, and is based purely on morphological parameters \cite{ChengIEEE2009}, wing kinematics \cite{ChangPNAS2014}, and the quasi-steady aerodynamic model. (c) A histogram of latency times across our data set, with latency time defined as the time between the onset of the magnetic perturbation and the beginning of a measurable corrective wing response ($\pm$ 4$\degree$ change in front stroke angle). The orange background in corresponds to the mean delay time $\pm$ standard deviation obtained from our controller model fits. (d) The time for the pitch correction--defined as the time it takes for the fly to recover 90\% of its original pitch orientation--plotted as a function of the maximum pitch deflection for each perturbation event. Solid and dashed lines give mean $\pm$ standard deviation. The lack of discernible correlation in d) is a hallmark of linear control.}
		\label{fig:stats}
\end{figure}

We also analyze the pitch correction timescales across our data set. Figure \ref{fig:stats}c shows a histogram of latency times for each corrective maneuver, defined as the time between the onset of the perturbation and the first measurable change in the front stroke angle ($|\Delta\phi_\text{w}^\text{front}| > 4\degree$). The mean latency time is 9.9 $\pm$ 2.1 ms corresponding to $\sim 2$ wing-beats (mean $\pm$ standard deviation, $n = 18$). Figure \ref{fig:stats}d plots the total correction time for each maneuver as a function of the maximum body pitch deflection in each perturbation event. We define the correction time as the time between the onset of the perturbation and the fly's correcting 90\% of the pitch deflection. The median correction time is 29 $\pm$ 8 ms (mean $\pm$ standard deviation, n = 32). Finally, we find that the correction time is weakly correlated with the perturbation amplitude (linear R$^2$ = 0.093), which is consistent with a linear control model.

\subsection{Control-Theory Model}
\label{subsec:ctrl}

\begin{figure}[h]
	\centering
		\includegraphics[scale = .9]{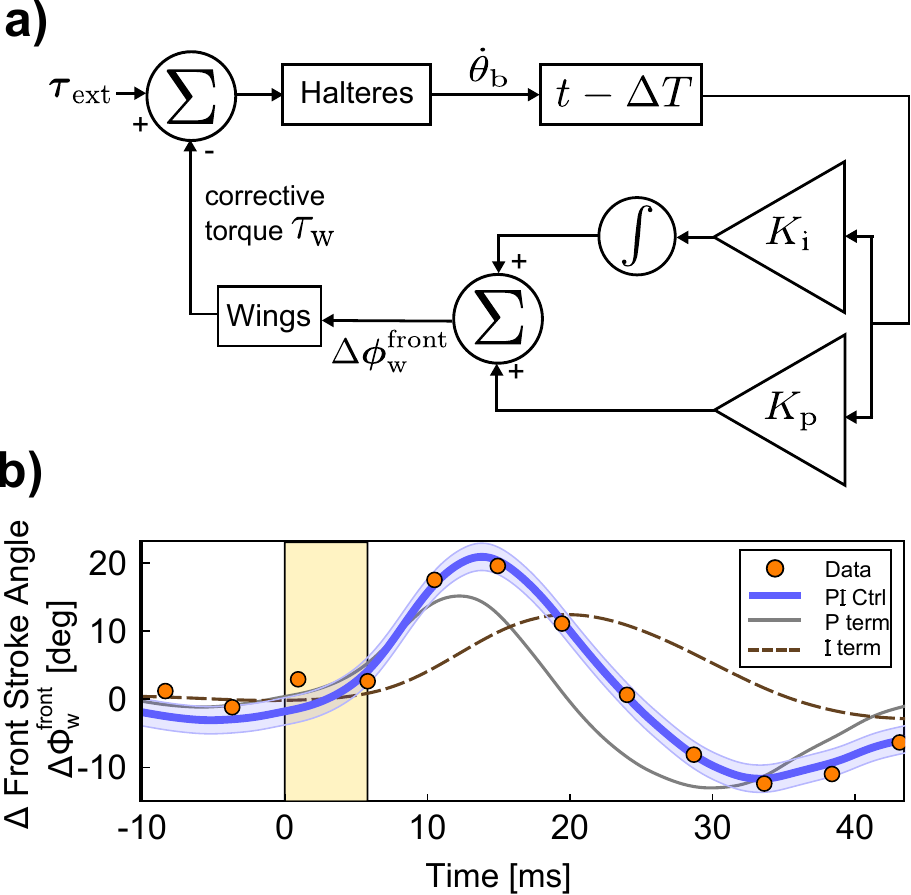}
		\caption{Control theory model. (a) A block diagram for the PI controller model. The effect of external torque ($\tau_\text{ext}$) is sensed by the halteres as angular velocity. The measured body pitch velocity ($\dot{\theta}_\text{b}$) is subject to a time delay ($\Delta T$), after which the signal is split into two branches. One branch is multiplied by $K_\text{i}$ and integrated to yield pitch displacement, while the other branch is multiplied by $K_\text{p}$. These signals are recombined as an output, $\Delta \phi^\text{front}_\text{w}$, that adjusts the front stroke angle of the wings. This adjustment to front stroke angle results in a corrective wing torque ($\tau_\text{w}$), which is in turn sensed by the fly. (b) Measured front stroke angle as a function of time for the pitch up event in Figures \ref{fig:bodykinematics}, \ref{fig:wingkinematics}, and \ref{fig:torquecomp} (orange dots) compared with the output of the fitted PI controller model (blue line). The relative contributions from the P and I terms are shown in the gray solid line and the brown dashed line, respectively. Shaded blue region corresponds to the confidence interval of the three control parameters. }
	 \label{fig:controller}
\end{figure}

We use a control-theoretic framework to describe flies' method for pitch stabilization. In particular, we model actuated changes to the front stroke angle as the output of a proportional-integral (PI) controller with time delay $\Delta T$, for which the input is body pitch velocity (block diagram in Figure \ref{fig:controller}a). The response $\Delta\phi_\text{w}^\text{front}$ is given by:
\begin{equation} \label{eq:ctrl}
\Delta\phi_\text{w}^\text{front}(t) = K_\text{p}\dot{\theta}_\text{b}(t - \Delta T) + K_\text{i}\Delta\theta_\text{b}(t- \Delta T)
\end{equation}
Equation \ref{eq:ctrl} states that adjustment of the front stroke angle ($\Delta\phi_\text{w}^\text{front}$) at a given time $t$ is given by a linear combination of the body pitch angle deviation from non-maneuvering orientation ($\Delta\theta_\text{b}$) and body pitch velocity ($\dot{\theta}_\text{b}$) at an earlier time $t - \Delta T$. The parameters $K_\text{p}$ and $ K_\text{i}$ are the proportionality constants that determine the relative weights of body pitch angle and pitch velocity. The same controller could be termed a PD controller, if the input to the controller were the body pitch angle. Because the fly halteres are known to measure body angular velocities \cite{PringleGyroscopic1948,DickinsonHaltere1999,NalbachNonOrthogonal1994}, we choose the PI nomenclature. We exclude controller models that depend on angular acceleration (PID) based on previous studies that have shown flies' corrective pitch response to be insensitive to angular acceleration \cite{DickinsonHaltere1999}. Further analysis of the data in \cite{DickinsonHaltere1999} confirms this model framework (see Supplementary). 

Importantly, a PI controller model with only angular velocity as an input cannot account for pitch stabilization on long timescales, due to integration errors affecting measurement of the absolute pitch angle. Controlling pitch on long timescales requires direct measurement of the pitch angle, as could be achieved by the visual system. An interplay between the haltere and visual systems, as in \cite{Huston2009Gated}, is necessary for comprehensive pitch stability. The PI model presented here can accurately account for the fly's fast reflex response, which stabilizes it against rapid pitch perturbations. 

Using measured values for $\Delta\phi_\text{w}^\text{front}$, $\Delta\theta_\text{b}$, and $\dot{\theta}_\text{b}$, we fit for the parameters $K_\text{p}$, $K_\text{i}$, and $\Delta T$. The three parameters are fit for each movie individually, using one data point per wing-stroke. The fit is performed by evaluating Equation \ref{eq:ctrl} on a dense 3D grid in parameter space and finding the global minimum for the sum of squared residuals between the control model and real data. The results of a controller fit for the pitch up event in Figures \ref{fig:bodykinematics}, \ref{fig:wingkinematics}, and \ref{fig:torquecomp} are shown in Figure \ref{fig:controller}b. The orange dots give the measured values of the front stroke angle, while the blue line shows the output of the fitted controller model. We find excellent quantitative agreement between the controller fit and our measured data with an RMS error of 1.9\degree, which is on the order of the measurement uncertainty for $\Delta\phi_\text{w}^\text{front}$. The controller model not only captures the sharp rise in $\Delta\phi_\text{w}^\text{front}$ in response to the perturbation, but also the subsequent decrease in $\Delta\phi_\text{w}^\text{front}$ corresponding to the braking counter-torque that slows the fly's downward pitching motion (Section \ref{subsec:aero}). The proportional (P) and integral (I) contributions to the controller model are shown in solid gray and dashed brown curves, respectively. The fast rise time of the response can be attributed to the proportional term.

We apply the same fitting process to nine movies. We find the values of fitted control parameters (Table I) to be $K_\text{i}$ = 0.3 $\pm$ 0.15, $K_\text{p}$ = 7 $\pm$ 2.1 ms, and $\Delta T$ = 6 $\pm$ 1.7 ms (mean $\pm$ standard deviation), with an average RMS fitting error of 3.0\degree. The mean value of $\Delta T$ corresponds to roughly 1 wing-beat, providing a lower bound for measured latency times (see Discussion). Figure \ref{fig:stats}c shows the region corresponding to mean $\Delta T\pm 1\sigma$ (highlighted in orange) compared with measured latency times. Confidence intervals (CI) in Table I are calculated based on a $\chi^2$ test for the fitting residuals in control parameter space. The confidence interval size is large relative to the fitted control parameters ($>50\%$ in some cases). The large confidence intervals, combined with the accuracy of the fit, indicate that the model is robust to deviations in the controller parameters. 

\begin{table}[h]
	\centering
	\begin{tabular}{|l|c|c|c|c|c|} 
		\hline
		Mov.				 	& $\Delta\theta_\text{b}$	& $K_\text{i}$ $\pm$ CI	&    $K_\text{p}$ $\pm$ CI & $\Delta T$ $\pm$ CI & RMS Err. \\
		Num.				 	& 				[deg]						&					[none]				&     [ms]								 & 		 [ms]            &    [deg] \\
		\hline \hline
		1 						&   				25						& 0.5 $\pm$ 0.24				& 	6 $\pm$ 1.9						 &	4 $\pm$ 2.1	    	 &		1.9					 \\	  
		\hline 
		2 						&   				16						& 0.1 $\pm$ 0.25				& 			8 $\pm$	3.4 		   &	8 $\pm$	2.5		     &		2.9		 		    \\	 
		\hline
		3 						&  					25						& 0.5 $\pm$ 0.19		& 			4 $\pm$ 1.7				&	4 $\pm$ 1.7			&		3.2						\\	 
		\hline
		4 						&  					6							& 0.5 $\pm$ 0.23		& 			12 $\pm$ 7				& 4 $\pm$ 3.6			&		1.7						\\	 
		\hline
		5 						& 					7							& 0.2 $\pm$ 0.61		& 		7 $\pm$ 7.3				  &	7 $\pm$ 5.8			&		2.5						\\	 
		\hline
		6 						& 					15						& 0.3 $\pm$ 0.24		& 		8 $\pm$ 3.2					&	6 $\pm$ 2.2			&		2.7						\\	 
		\hline
		7 						& 					-24						& 0.3 $\pm$ 0.29		& 		6 $\pm$ 2.4					&	7 $\pm$ 2.1			&		3.7							\\	 
		\hline
		8 						& 					-23						& 0.2 $\pm$ 0.31		& 		6 $\pm$ 2.0					&	9 $\pm$ 1.5			&		2.9						\\	 
		\hline
		9 						& 					-21						& 0.2 $\pm$ 0.26		& 		8 $\pm$ 2.8					&	7 $\pm$ 3.6			&		4.5					\\	 
		\hline
	\end{tabular}
	\label{tab1}
	\caption{Fit results for PI controller model with confidence intervals (CI) for each parameter.}
\end{table}

\subsection{Numerical Simulation}
\label{subsec:sim}
\begin{figure}[h]
	\centering
		\includegraphics[scale = .85]{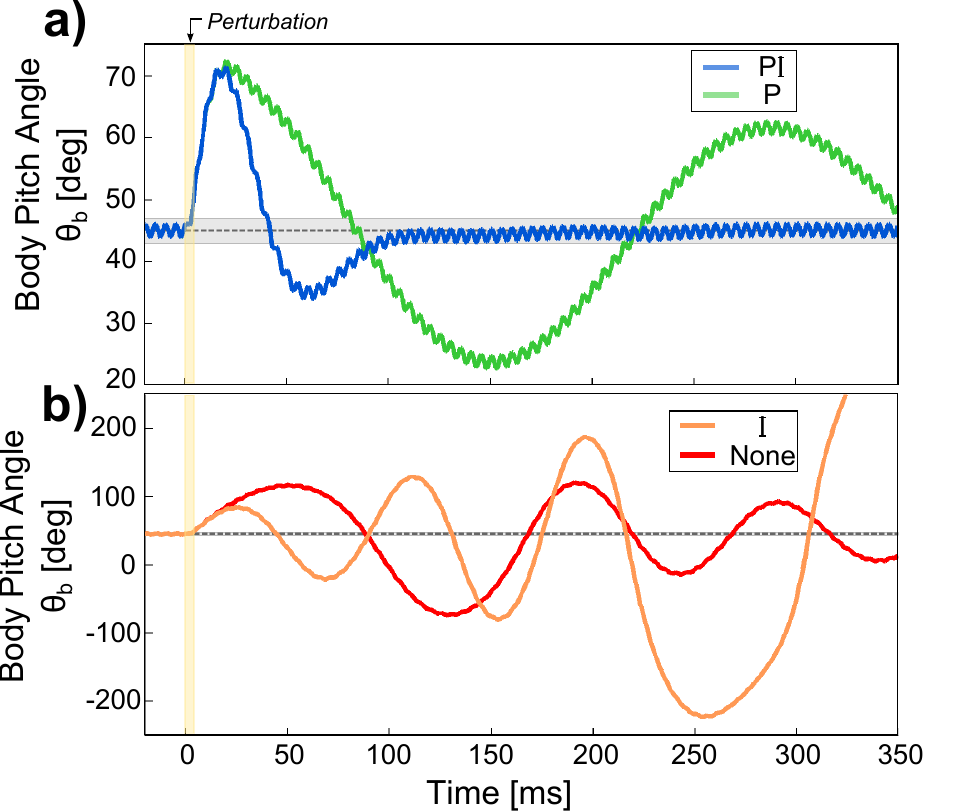}
	  \caption{Numerical simulation results for different controller models. (a) Time series of body pitch angle for simulated flies implementing proportional-integral (PI, blue) and proportional (P, green) control. (b) Time series of body pitch angle for simulated flies implementing integral (I, orange) and no (None, red) control. Time axes for both (a) and (b) are the same, but the range of the pitch angle axis differs significantly between the two. Gray region in (a) and (b) corresponds to 45 $\pm$ 2\degree, where 45$\degree$ is the reference body pitch angle for each controller. Hence, curves returning to and remaining within this region indicate successful control. The yellow strip indicates the duration of the mechanical perturbation (0 - 5 ms).}
	 \label{fig:sim}
\end{figure}

To corroborate our experimental evidence for the PI controller, we perform a dynamical simulation of a mechanically perturbed fruit fly. The simulation solves the equations of motion for the pitch, longitudinal, and vertical degrees of freedom, assuming the fly's geometry, simplified wing kinematics, and the quasi-steady aerodynamic force model detailed above. The body pitch angle over time for simulated flies implementing different control strategies is shown in Figure \ref{fig:sim}. The four simulated control schemes shown are: i) proportional-integral (blue), ii) proportional (green), iii) integral control (orange), and iv) no control (red). To determine parameters for the simulated controllers, we fit our experimental data to each model, and to mimic the perturbation conditions in our experiments we impose a 5 ms external mechanical torque on the simulated flies (yellow strip), with magnitude comparable to our real system. For details on the simulation, see Supplementary.

We find that flies with no control or I control are subject to large, rapid oscillations of body pitch angle (Figure \ref{fig:sim}b), while flies with P control are subject to slightly smaller, long-timescale oscillations (Figure \ref{fig:sim}a). In all three of these cases, the simulated fly fails to remain aloft and rapidly loses altitude. Among the four candidate models, PI control is the only one that is consistent with the fast, robust pitch control that we observe experimentally (Figure \ref{fig:sim}a). Simulated flies implementing PI control correct their orientation over timescales similar to those in our experimental data. In contrast to the other three controller models, the simulated flies using PI control maintain pitch stability over long times and remain aloft. The general features of each control scheme show little sensitivity to the values of the control parameters, in agreement with the large confidence intervals that we find for fitted control parameters in Section \ref{subsec:ctrl}.

\subsection{Extreme Perturbations}

\begin{figure}[h]
		\includegraphics[scale = .8]{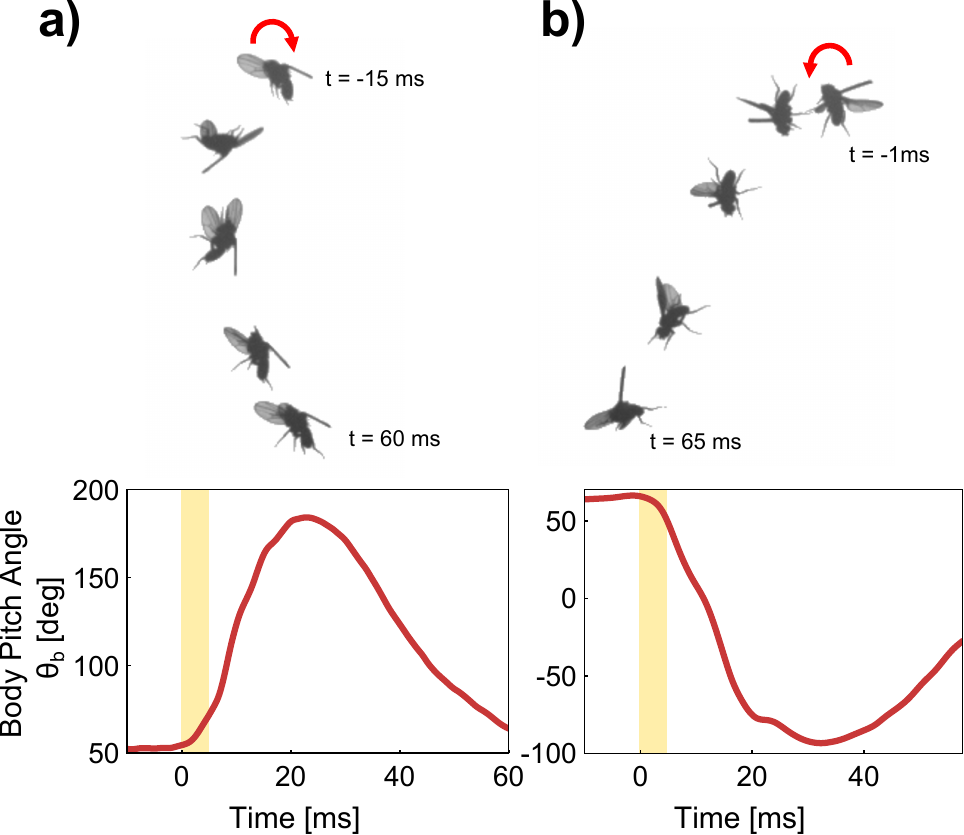}
	  \caption{Large perturbations. Overlaid snapshots from raw data and time series of body pitch angle for (a) pitch up perturbation and (b) pitch down perturbation (Movie S2). The pitched up fly reaches a maximum pitch deflection of 130$\degree$ at $\sim$20 ms after the onset of the perturbation. The pitched down fly reaches a maximum pitch deflection of -155$\degree$ after $\sim$30 ms. The loss of altitude during the correction is evident in both cases and shown to scale. Yellow strips indicate the 5.8 ms magnetic pulse; red arrows indicate the direction of each perturbation.}
	 \label{fig:extreme}
\end{figure}

In addition to the 18 perturbation events analyzed in full (Figure \ref{fig:bodykinematics}, \ref{fig:stats}), we examine two large-amplitude perturbation events. Snapshots and time courses of body pitch angle are shown in Figure \ref{fig:extreme} for a pitch up and a pitch down event, both with maximum pitch deflection greater than 130\degree (Movie S2). Remarkably, both flies performed successful correction maneuvers, although they were not in-frame long enough to observe them returning to their original orientation. The correction time for both large-amplitude events ($>50$ ms) is longer than the correction times shown in Figure \ref{fig:stats}, which can be attributed to the fact that the controlled quantity $\Delta\phi_\text{w}^\text{front}$ is biologically constrained: front stroke angle is limited, for instance, by the angle at which the body or the other wing obstructs a wing's forward motion. If we input the body pitch kinematics for the events in Figure \ref{fig:extreme} into our PI controller model, the model predicts changes to $\phi_\text{w}^\text{front}$ in excess of 100\degree--a value that is physiologically impossible in the forward direction and not observed in the backward direction. 
Assuming the flies' corrective response is bounded by $|\Delta\phi_\text{w}^\text{front}| \le 30\degree$, a physiologically reasonable estimate, our numerical simulation predicts a response time of $\sim$70 ms for a perturbation $\Delta\theta_\text{b} = -150\degree$, in excellent agreement with the experimental data.

\section{Discussion}
\label{sec:discussion}

\subsection{Front Stroke Angle as the Controlled Quantity}

We show that front stroke angle modulation is the primary mechanism for body pitch control in fruit flies, consistent with previous experiments \cite{ZankerIIIControl1990, DickinsonHaltere1999, RistrophPitchInterface2013} and is in the same spirit as other proposed mechanisms that include modulation of the mid-stroke angle \cite{ChangPNAS2014}. Our computational results (Figures \ref{fig:stats} and \ref{fig:sim}) show that a minimal model, which only incorporates changes to front stroke angle, and uses control parameters extracted from fits to our measurements, is the simplest linear, continuous model capable of stabilizing the body pitch angle on time scales similar to those observed in the experiments. 

Kinematic variables other than $\phi_\text{w}^\text{front}$ may also contribute to pitch correction. Previous studies have associated changes to stroke plane deviation \cite{ZankerLongitudinal1988}, wing-beat frequency \cite{DickinsonHaltere1999}, and body posture \cite{TaylorMechanics2001} with pitch correction. In particular, we observe transient alterations in both wing pitch and deviation angle during corrective maneuvers. Figure \ref{fig:leifLike} suggests that, when combined with modulation of front stroke angle, changes to wing pitch and deviation angle can account for up to 40\% of body pitch correction, consistent with \cite{MuijresScience2014}. The detailed role of these kinematic variables in pitch control and whether they are actively or passively actuated remains unknown.

\subsection{Discrete vs. Continuous Control Models}

The periodic motion of wing flapping introduces inherent discreteness to insect flight. For processes occurring on timescales comparable to a wing-stroke period--like the perturbations and maneuvers we reported here--we expect discrete effects to be more pronounced. In particular, modulations of front stroke angle can, by definition, occur only once per wing-beat. Because perturbations can be induced at any time during the wing-beat, but the actuated kinematics are discretely constrained, latency times for correction depend on the phase of the perturbation relative to the wing-stroke. Latency times will be bounded from below by the flies' neural response time, but could potentially be as much as one wing-beat longer as a result of the phase of the perturbation within the wing-beat. 

Measured latency times can also depend on discrete sensing. The temporal sampling resolution with which flies can measure mechanical perturbations is likely determined by motion of their halteres, the rate-gyro sensory organs used in fast perturbation response \cite{DickinsonHaltere1999}. Dipteran halteres beat at the wing frequency and use Coriolis forces to measure body angular velocities \cite{PringleExcitation1949, NalbachNonOrthogonal1994}. The largest sensitivity to mechanical perturbations is likely to occur at times during the fly's mid-stroke, when Coriolis forces on the halteres are the largest \cite{NalbachNonOrthogonal1994}. Sensing at discrete times introduces a second relevant phase for correction latency time: the phase of the perturbation relative to sensing. Similar to discrete actuation, discrete sensing would lead to latency times longer than the neural response time. Moreover, even during the fly's mid-stroke, its halteres only have finite sensitivity. It is likely that there exists some threshold for angular velocities that are large enough to elicit a control response \cite{FoxHaltere2008}.

The continuous PI controller model does not account for the effects of discrete actuation, discrete sampling, or sensing threshold. Hence, the measured latency time should constitute an upper bound for the delay time that we obtain from the controller model. Indeed, the time delays from our controller model ($\Delta T =  6 \pm 1.7$  ms) is roughly one wing-beat shorter than the measured latency times ($9.9 \pm 2.1$ ms). 

Despite the inherent discreteness of the fly control systems, our continuous PI controller model quantitatively captures the behavior of flies in response to pitch perturbations (Figure \ref{fig:controller}). This quantitative agreement leads to an interesting open question: under what conditions does it become necessary to use a discrete controller model to describe flight stabilization? To answer this question would require precise perturbation timing, in order to probe the short timescales at which discretization becomes relevant. Such an analysis could provide significant insight into the timing and thresholding of fruit fly reflexes.
 
\subsection{Physiological Basis for Pitch Control}

Both the mechanism and timing for the pitch correction indicate a likely candidate muscle for control actuation: the first basalare muscle (b1), as suggested by previous studies \cite{FayyazuddinMuscle1999, ChangPNAS2014}. Among dipteran flight control muscles, b1 is unique in that it is active during every wing-stroke \cite{HeideNeural1983, HeideOptomotor1996, MiyanDiptera1985}, which would allow for the wing-beat timescale pitch control that we observe. Moreover, b1 activity is strongly correlated with modulations of ventral stroke amplitude, i.e. changes in $\phi_\text{w}^\text{front}$ \cite{TuDickinsonMuscle1997, WalkerMicrotomography2014}. In blowflies, b1 activity is also correlated with changes in deviation angle during the ventral stroke \cite{BalintDickinsonJEB2004}, which could explain the slight shifts in deviation angle that we observe during correction (Figure \ref{fig:wingkinematics}b,f).  Our results indirectly support the hypothesis that the b1 muscle is responsible for pitch control through the regulation of $\phi_\text{w}^\text{front}$. Testing flies with disabled or altered b1 muscles could provide an avenue for confirming the role of b1 in the pitch control process.

\subsection{Linear Control of Body Orientation}
In addition to the results on fruit flies reported here, PI control has also been identified in pitch control for hawkmoths \cite{ChengHedrickJEB2011, WindsorInterface2014}. The anatomical similarities found across species suggest that pitch instability is an obstacle faced by many flapping insects \cite{SunDynamicStability2007,RistrophPitchInterface2013}; a natural question raised by these collective findings is whether or not PI control is a generic feature of pitch stabilization in insects. Beyond insects, what we refer to as PI control has also been observed in fast obstacle avoidance in pigeons \cite{lin2014through}. Future research on the ubiquity of PI control could have fascinating implications for the evolution of flight stabilization mechanisms.

Extending beyond pitch stabilization, our results, together with previous studies on yaw \cite{RistrophPNAS2010} and roll \cite{BeatusInterface2015} control in fruit flies, show that the strategies flies use to control each of their body Euler angles can be modeled as PI controllers. However, the overarching structure in which these three individual controllers are embedded is still unknown. Given the non-commutativity of rotations in 3D, the relationship between controllers that measure different angular coordinates is likely to be non-trivial. For example, in response to perturbations that simultaneously affect both the roll and yaw degrees of freedom, previous studies observed preferential correction for roll over yaw \cite{BeatusInterface2015}. Because roll is known to be aerodynamically unstable, while yaw is passively stable \cite{ZhangSunLateral2010, GaoCFD2011, SunRevModPhys2014}, the results from \cite{BeatusInterface2015} suggest that control for different body angles may be imposed hierarchically, with preference given to correcting degrees of freedom that are most unstable. However, in our data set we observe perturbation events in which pitch is corrected prior to roll, despite roll being the more unstable degree of freedom, implying that the response to complex perturbations has amplitude dependent features. Taken together, these results hint at a complex and intriguing control architecture used by flies to stabilize their orientation.

An understanding of the relationship between control of different Euler angles could have profound implications for how the fly encodes information about its body orientation. In the case of vision, organism-specific demands have spurred the development of novel, specialized neural structures in both mammals \cite{Hafting2005Microstructure, Yartsev2011Grid} and insects \cite{Ofstad2011Visual, Seelig2013Feature}.  Pioneering work on information processing from halteres has suggested similar morphology/function relationships for the gyroscopic rate sensing in insects \cite{Fox2010Encoding}. Connecting such analyses with the resultant control structure observed in free-flight behavioral experiments could provide a window into the most basic ways in which flies sense and interpret the world. 

\section*{Acknowledgments}
\vspace{-3mm}
We thank Grace (Li) Chi and Andy Clark for providing flies;  Ty Hedrick for advice on camera calibration; Andy Ruina and the Cohen group for useful conversations. 
\vspace{-3mm}
\section*{Funding}
\vspace{-3mm}
This work was supported in part by an NSF DMR award (no. 1056662) and in part by an ARO award (no. 61651-EG). S.C.W. was supported by the NDSEG Fellowship. T.B. was supported by the Cross Disciplinary Post-Doctoral Fellowship of the Human Frontier Science Program. 

\bibliography{FlySources}

\begin{thebibliography}{61}
\expandafter\ifx\csname natexlab\endcsname\relax\def\natexlab#1{#1}\fi
\expandafter\ifx\csname bibnamefont\endcsname\relax
  \def\bibnamefont#1{#1}\fi
\expandafter\ifx\csname bibfnamefont\endcsname\relax
  \def\bibfnamefont#1{#1}\fi
\expandafter\ifx\csname citenamefont\endcsname\relax
  \def\citenamefont#1{#1}\fi
\expandafter\ifx\csname url\endcsname\relax
  \def\url#1{\texttt{#1}}\fi
\expandafter\ifx\csname urlprefix\endcsname\relax\def\urlprefix{URL }\fi
\providecommand{\bibinfo}[2]{#2}
\providecommand{\eprint}[2][]{\url{#2}}

\bibitem[{\citenamefont{Taylor and Thomas}(2003)}]{TaylorThomasJEB2003}
\bibinfo{author}{\bibfnamefont{G.~K.} \bibnamefont{Taylor}} \bibnamefont{and}
  \bibinfo{author}{\bibfnamefont{A.~L.} \bibnamefont{Thomas}},
  \bibinfo{journal}{Journal of Experimental Biology}
  \textbf{\bibinfo{volume}{206}}, \bibinfo{pages}{2803} (\bibinfo{year}{2003}).

\bibitem[{\citenamefont{Sun and Xiong}(2005)}]{SunBumblebeeJEB2005}
\bibinfo{author}{\bibfnamefont{M.}~\bibnamefont{Sun}} \bibnamefont{and}
  \bibinfo{author}{\bibfnamefont{Y.}~\bibnamefont{Xiong}},
  \bibinfo{journal}{The Journal of experimental biology}
  \textbf{\bibinfo{volume}{208}}, \bibinfo{pages}{447} (\bibinfo{year}{2005}).

\bibitem[{\citenamefont{Taylor and {\.Z}bikowski}(2005)}]{TaylorZibkowski2005}
\bibinfo{author}{\bibfnamefont{G.~K.} \bibnamefont{Taylor}} \bibnamefont{and}
  \bibinfo{author}{\bibfnamefont{R.}~\bibnamefont{{\.Z}bikowski}},
  \bibinfo{journal}{Journal of The Royal Society Interface}
  \textbf{\bibinfo{volume}{2}}, \bibinfo{pages}{197} (\bibinfo{year}{2005}).

\bibitem[{\citenamefont{Sun et~al.}(2007)\citenamefont{Sun, Wang, and
  Xiong}}]{SunDynamicStability2007}
\bibinfo{author}{\bibfnamefont{M.}~\bibnamefont{Sun}},
  \bibinfo{author}{\bibfnamefont{J.}~\bibnamefont{Wang}}, \bibnamefont{and}
  \bibinfo{author}{\bibfnamefont{Y.}~\bibnamefont{Xiong}},
  \bibinfo{journal}{Acta Mechanica Sinica} \textbf{\bibinfo{volume}{23}},
  \bibinfo{pages}{231} (\bibinfo{year}{2007}).

\bibitem[{\citenamefont{Liu et~al.}(2010)\citenamefont{Liu, Nakata, Gao, Maeda,
  Aono, and Shyy}}]{LiuCFD2010}
\bibinfo{author}{\bibfnamefont{H.}~\bibnamefont{Liu}},
  \bibinfo{author}{\bibfnamefont{T.}~\bibnamefont{Nakata}},
  \bibinfo{author}{\bibfnamefont{N.}~\bibnamefont{Gao}},
  \bibinfo{author}{\bibfnamefont{M.}~\bibnamefont{Maeda}},
  \bibinfo{author}{\bibfnamefont{H.}~\bibnamefont{Aono}}, \bibnamefont{and}
  \bibinfo{author}{\bibfnamefont{W.}~\bibnamefont{Shyy}},
  \bibinfo{journal}{Acta Mechanica Sinica} \textbf{\bibinfo{volume}{26}},
  \bibinfo{pages}{863} (\bibinfo{year}{2010}).

\bibitem[{\citenamefont{Faruque and Humbert}(2010)}]{FaruqueHumbert2010a}
\bibinfo{author}{\bibfnamefont{I.}~\bibnamefont{Faruque}} \bibnamefont{and}
  \bibinfo{author}{\bibfnamefont{J.~S.} \bibnamefont{Humbert}},
  \bibinfo{journal}{Journal of theoretical biology}
  \textbf{\bibinfo{volume}{264}}, \bibinfo{pages}{538} (\bibinfo{year}{2010}).

\bibitem[{\citenamefont{Zhang and Sun}(2010)}]{ZhangSunLateral2010}
\bibinfo{author}{\bibfnamefont{Y.}~\bibnamefont{Zhang}} \bibnamefont{and}
  \bibinfo{author}{\bibfnamefont{M.}~\bibnamefont{Sun}}, \bibinfo{journal}{Acta
  Mechanica Sinica} \textbf{\bibinfo{volume}{26}}, \bibinfo{pages}{175}
  (\bibinfo{year}{2010}).

\bibitem[{\citenamefont{Zhang and Sun}(2011)}]{ZhangSunLateralControl2011}
\bibinfo{author}{\bibfnamefont{Y.-L.} \bibnamefont{Zhang}} \bibnamefont{and}
  \bibinfo{author}{\bibfnamefont{M.}~\bibnamefont{Sun}}, \bibinfo{journal}{Acta
  Mechanica Sinica} \textbf{\bibinfo{volume}{27}}, \bibinfo{pages}{823}
  (\bibinfo{year}{2011}).

\bibitem[{\citenamefont{Gao et~al.}(2011)\citenamefont{Gao, Aono, and
  Liu}}]{GaoCFD2011}
\bibinfo{author}{\bibfnamefont{N.}~\bibnamefont{Gao}},
  \bibinfo{author}{\bibfnamefont{H.}~\bibnamefont{Aono}}, \bibnamefont{and}
  \bibinfo{author}{\bibfnamefont{H.}~\bibnamefont{Liu}},
  \bibinfo{journal}{Journal of Theoretical Biology}
  \textbf{\bibinfo{volume}{270}}, \bibinfo{pages}{98} (\bibinfo{year}{2011}).

\bibitem[{\citenamefont{Pérez-Arancibia
  et~al.}(2011)\citenamefont{Pérez-Arancibia, Ma, Galloway, Greenberg, and
  Wood}}]{WoodVerticalFlight2011}
\bibinfo{author}{\bibfnamefont{N.~O.} \bibnamefont{Pérez-Arancibia}},
  \bibinfo{author}{\bibfnamefont{K.~Y.} \bibnamefont{Ma}},
  \bibinfo{author}{\bibfnamefont{K.~C.} \bibnamefont{Galloway}},
  \bibinfo{author}{\bibfnamefont{J.~D.} \bibnamefont{Greenberg}},
  \bibnamefont{and} \bibinfo{author}{\bibfnamefont{R.~J.} \bibnamefont{Wood}},
  \bibinfo{journal}{Bioinspiration \& Biomimetics}
  \textbf{\bibinfo{volume}{6}}, \bibinfo{pages}{036009} (\bibinfo{year}{2011}).

\bibitem[{\citenamefont{Ristroph et~al.}(2013)\citenamefont{Ristroph, Ristroph,
  Morozova, Bergou, Chang, Guckenheimer, Wang, and
  Cohen}}]{RistrophPitchInterface2013}
\bibinfo{author}{\bibfnamefont{L.}~\bibnamefont{Ristroph}},
  \bibinfo{author}{\bibfnamefont{G.}~\bibnamefont{Ristroph}},
  \bibinfo{author}{\bibfnamefont{S.}~\bibnamefont{Morozova}},
  \bibinfo{author}{\bibfnamefont{A.~J.} \bibnamefont{Bergou}},
  \bibinfo{author}{\bibfnamefont{S.}~\bibnamefont{Chang}},
  \bibinfo{author}{\bibfnamefont{J.}~\bibnamefont{Guckenheimer}},
  \bibinfo{author}{\bibfnamefont{Z.~J.} \bibnamefont{Wang}}, \bibnamefont{and}
  \bibinfo{author}{\bibfnamefont{I.}~\bibnamefont{Cohen}},
  \bibinfo{journal}{Journal of The Royal Society Interface}
  \textbf{\bibinfo{volume}{10}} (\bibinfo{year}{2013}).

\bibitem[{\citenamefont{Xu and Sun}(2013)}]{XuSunCFD2013}
\bibinfo{author}{\bibfnamefont{N.}~\bibnamefont{Xu}} \bibnamefont{and}
  \bibinfo{author}{\bibfnamefont{M.}~\bibnamefont{Sun}},
  \bibinfo{journal}{Journal of theoretical biology}
  \textbf{\bibinfo{volume}{319}}, \bibinfo{pages}{102} (\bibinfo{year}{2013}).

\bibitem[{\citenamefont{Sun}(2014)}]{SunRevModPhys2014}
\bibinfo{author}{\bibfnamefont{M.}~\bibnamefont{Sun}}, \bibinfo{journal}{Rev.
  Mod. Phys.} \textbf{\bibinfo{volume}{86}}, \bibinfo{pages}{615}
  (\bibinfo{year}{2014}).

\bibitem[{\citenamefont{Beatus et~al.}(2015)\citenamefont{Beatus, Guckenheimer,
  and Cohen}}]{BeatusInterface2015}
\bibinfo{author}{\bibfnamefont{T.}~\bibnamefont{Beatus}},
  \bibinfo{author}{\bibfnamefont{J.~M.} \bibnamefont{Guckenheimer}},
  \bibnamefont{and} \bibinfo{author}{\bibfnamefont{I.}~\bibnamefont{Cohen}},
  \bibinfo{journal}{Journal of The Royal Society Interface}
  \textbf{\bibinfo{volume}{12}} (\bibinfo{year}{2015}), ISSN
  \bibinfo{issn}{1742-5689}.

\bibitem[{\citenamefont{Combes and Dudley}(2009)}]{CombesPNAS2009}
\bibinfo{author}{\bibfnamefont{S.~A.} \bibnamefont{Combes}} \bibnamefont{and}
  \bibinfo{author}{\bibfnamefont{R.}~\bibnamefont{Dudley}},
  \bibinfo{journal}{Proceedings of the National Academy of Sciences}
  \textbf{\bibinfo{volume}{106}}, \bibinfo{pages}{9105} (\bibinfo{year}{2009}).

\bibitem[{\citenamefont{Dickerson et~al.}(2012)\citenamefont{Dickerson,
  Shankles, Madhavan, and Hu}}]{DickersonRaindrop2012}
\bibinfo{author}{\bibfnamefont{A.~K.} \bibnamefont{Dickerson}},
  \bibinfo{author}{\bibfnamefont{P.~G.} \bibnamefont{Shankles}},
  \bibinfo{author}{\bibfnamefont{N.~M.} \bibnamefont{Madhavan}},
  \bibnamefont{and} \bibinfo{author}{\bibfnamefont{D.~L.} \bibnamefont{Hu}},
  \bibinfo{journal}{Proceedings of the National Academy of Sciences}
  \textbf{\bibinfo{volume}{109}}, \bibinfo{pages}{9822} (\bibinfo{year}{2012}).

\bibitem[{\citenamefont{Ravi et~al.}(2013)\citenamefont{Ravi, Crall, Fisher,
  and Combes}}]{RaviJEB2013}
\bibinfo{author}{\bibfnamefont{S.}~\bibnamefont{Ravi}},
  \bibinfo{author}{\bibfnamefont{J.~D.} \bibnamefont{Crall}},
  \bibinfo{author}{\bibfnamefont{A.}~\bibnamefont{Fisher}}, \bibnamefont{and}
  \bibinfo{author}{\bibfnamefont{S.~A.} \bibnamefont{Combes}},
  \bibinfo{journal}{The Journal of Experimental Biology}
  \textbf{\bibinfo{volume}{216}}, \bibinfo{pages}{4299} (\bibinfo{year}{2013}).

\bibitem[{\citenamefont{Ortega-Jimenez
  et~al.}(2013)\citenamefont{Ortega-Jimenez, Greeter, Mittal, and
  Hedrick}}]{OrtegaJEB2013}
\bibinfo{author}{\bibfnamefont{V.~M.} \bibnamefont{Ortega-Jimenez}},
  \bibinfo{author}{\bibfnamefont{J.~S.} \bibnamefont{Greeter}},
  \bibinfo{author}{\bibfnamefont{R.}~\bibnamefont{Mittal}}, \bibnamefont{and}
  \bibinfo{author}{\bibfnamefont{T.~L.} \bibnamefont{Hedrick}},
  \bibinfo{journal}{The Journal of experimental biology}
  \textbf{\bibinfo{volume}{216}}, \bibinfo{pages}{4567} (\bibinfo{year}{2013}).

\bibitem[{\citenamefont{Vance et~al.}(2013)\citenamefont{Vance, Faruque, and
  Humbert}}]{VanceGust2013}
\bibinfo{author}{\bibfnamefont{J.}~\bibnamefont{Vance}},
  \bibinfo{author}{\bibfnamefont{I.}~\bibnamefont{Faruque}}, \bibnamefont{and}
  \bibinfo{author}{\bibfnamefont{J.}~\bibnamefont{Humbert}},
  \bibinfo{journal}{Bioinspiration \& biomimetics}
  \textbf{\bibinfo{volume}{8}}, \bibinfo{pages}{016004} (\bibinfo{year}{2013}).

\bibitem[{\citenamefont{Chang and Wang}(2014)}]{ChangPNAS2014}
\bibinfo{author}{\bibfnamefont{S.}~\bibnamefont{Chang}} \bibnamefont{and}
  \bibinfo{author}{\bibfnamefont{Z.~J.} \bibnamefont{Wang}},
  \bibinfo{journal}{Proceedings of the National Academy of Sciences}
  \textbf{\bibinfo{volume}{111}}, \bibinfo{pages}{11246}
  (\bibinfo{year}{2014}).

\bibitem[{\citenamefont{Zanker}(1990)}]{ZankerIIIControl1990}
\bibinfo{author}{\bibfnamefont{J.}~\bibnamefont{Zanker}},
  \bibinfo{journal}{Philosophical Transactions of the Royal Society of London.
  B, Biological Sciences} \textbf{\bibinfo{volume}{327}}, \bibinfo{pages}{43}
  (\bibinfo{year}{1990}).

\bibitem[{\citenamefont{Nalbach}(1994)}]{NalbachNonOrthogonal1994}
\bibinfo{author}{\bibfnamefont{G.}~\bibnamefont{Nalbach}},
  \bibinfo{journal}{Neuroscience} \textbf{\bibinfo{volume}{61}},
  \bibinfo{pages}{149} (\bibinfo{year}{1994}).

\bibitem[{\citenamefont{Dickinson}(1999)}]{DickinsonHaltere1999}
\bibinfo{author}{\bibfnamefont{M.~H.} \bibnamefont{Dickinson}},
  \bibinfo{journal}{Philosophical Transactions of the Royal Society of
  London.Series B: Biological Sciences} \textbf{\bibinfo{volume}{354}},
  \bibinfo{pages}{903} (\bibinfo{year}{1999}).

\bibitem[{\citenamefont{Sherman and Dickinson}(2004)}]{ShermanDickinsonJEB2003}
\bibinfo{author}{\bibfnamefont{A.}~\bibnamefont{Sherman}} \bibnamefont{and}
  \bibinfo{author}{\bibfnamefont{M.~H.} \bibnamefont{Dickinson}},
  \bibinfo{journal}{Journal of experimental biology}
  \textbf{\bibinfo{volume}{207}}, \bibinfo{pages}{133} (\bibinfo{year}{2004}).

\bibitem[{\citenamefont{Sherman and Dickinson}(2003)}]{ShermanDickinsonJEB2004}
\bibinfo{author}{\bibfnamefont{A.}~\bibnamefont{Sherman}} \bibnamefont{and}
  \bibinfo{author}{\bibfnamefont{M.~H.} \bibnamefont{Dickinson}},
  \bibinfo{journal}{Journal of Experimental Biology}
  \textbf{\bibinfo{volume}{206}}, \bibinfo{pages}{295} (\bibinfo{year}{2003}).

\bibitem[{\citenamefont{Fry et~al.}(2005)\citenamefont{Fry, Sayaman, and
  Dickinson}}]{FryJEB2005}
\bibinfo{author}{\bibfnamefont{S.~N.} \bibnamefont{Fry}},
  \bibinfo{author}{\bibfnamefont{R.}~\bibnamefont{Sayaman}}, \bibnamefont{and}
  \bibinfo{author}{\bibfnamefont{M.~H.} \bibnamefont{Dickinson}},
  \bibinfo{journal}{Journal of Experimental Biology}
  \textbf{\bibinfo{volume}{208}}, \bibinfo{pages}{2303} (\bibinfo{year}{2005}).

\bibitem[{\citenamefont{Bender and Dickinson}(2006)}]{BenderVisual2006}
\bibinfo{author}{\bibfnamefont{J.~A.} \bibnamefont{Bender}} \bibnamefont{and}
  \bibinfo{author}{\bibfnamefont{M.~H.} \bibnamefont{Dickinson}},
  \bibinfo{journal}{The Journal of experimental biology}
  \textbf{\bibinfo{volume}{209}}, \bibinfo{pages}{3170} (\bibinfo{year}{2006}).

\bibitem[{\citenamefont{Ennos}(1989)}]{EnnosJEB1989}
\bibinfo{author}{\bibfnamefont{A.~R.} \bibnamefont{Ennos}},
  \bibinfo{journal}{Journal of Experimental Biology}
  \textbf{\bibinfo{volume}{142}}, \bibinfo{pages}{49} (\bibinfo{year}{1989}).

\bibitem[{\citenamefont{Fry et~al.}(2003)\citenamefont{Fry, Sayaman, and
  Dickinson}}]{FryScience2003}
\bibinfo{author}{\bibfnamefont{S.~N.} \bibnamefont{Fry}},
  \bibinfo{author}{\bibfnamefont{R.}~\bibnamefont{Sayaman}}, \bibnamefont{and}
  \bibinfo{author}{\bibfnamefont{M.~H.} \bibnamefont{Dickinson}},
  \bibinfo{journal}{Science} \textbf{\bibinfo{volume}{300}},
  \bibinfo{pages}{495} (\bibinfo{year}{2003}).

\bibitem[{\citenamefont{Ristroph et~al.}(2009)\citenamefont{Ristroph, Berman,
  Bergou, Wang, and Cohen}}]{RistrophJEB2009}
\bibinfo{author}{\bibfnamefont{L.}~\bibnamefont{Ristroph}},
  \bibinfo{author}{\bibfnamefont{G.~J.} \bibnamefont{Berman}},
  \bibinfo{author}{\bibfnamefont{A.~J.} \bibnamefont{Bergou}},
  \bibinfo{author}{\bibfnamefont{Z.~J.} \bibnamefont{Wang}}, \bibnamefont{and}
  \bibinfo{author}{\bibfnamefont{I.}~\bibnamefont{Cohen}},
  \bibinfo{journal}{Journal of Experimental Biology}
  \textbf{\bibinfo{volume}{212}}, \bibinfo{pages}{1324} (\bibinfo{year}{2009}).

\bibitem[{\citenamefont{Cheng et~al.}(2011)\citenamefont{Cheng, Deng, and
  Hedrick}}]{ChengHedrickJEB2011}
\bibinfo{author}{\bibfnamefont{B.}~\bibnamefont{Cheng}},
  \bibinfo{author}{\bibfnamefont{X.}~\bibnamefont{Deng}}, \bibnamefont{and}
  \bibinfo{author}{\bibfnamefont{T.~L.} \bibnamefont{Hedrick}},
  \bibinfo{journal}{The Journal of Experimental Biology}
  \textbf{\bibinfo{volume}{214}}, \bibinfo{pages}{4092} (\bibinfo{year}{2011}).

\bibitem[{\citenamefont{Windsor et~al.}(2014)\citenamefont{Windsor, Bomphrey,
  and Taylor}}]{WindsorInterface2014}
\bibinfo{author}{\bibfnamefont{S.~P.} \bibnamefont{Windsor}},
  \bibinfo{author}{\bibfnamefont{R.~J.} \bibnamefont{Bomphrey}},
  \bibnamefont{and} \bibinfo{author}{\bibfnamefont{G.~K.}
  \bibnamefont{Taylor}}, \bibinfo{journal}{Journal of The Royal Society
  Interface} \textbf{\bibinfo{volume}{11}}, \bibinfo{pages}{20130921}
  (\bibinfo{year}{2014}).

\bibitem[{\citenamefont{Muijres et~al.}(2014)\citenamefont{Muijres, Elzinga,
  Melis, and Dickinson}}]{MuijresScience2014}
\bibinfo{author}{\bibfnamefont{F.~T.} \bibnamefont{Muijres}},
  \bibinfo{author}{\bibfnamefont{M.~J.} \bibnamefont{Elzinga}},
  \bibinfo{author}{\bibfnamefont{J.~M.} \bibnamefont{Melis}}, \bibnamefont{and}
  \bibinfo{author}{\bibfnamefont{M.~H.} \bibnamefont{Dickinson}},
  \bibinfo{journal}{Science} \textbf{\bibinfo{volume}{344}},
  \bibinfo{pages}{172} (\bibinfo{year}{2014}).

\bibitem[{\citenamefont{Ristroph et~al.}(2010)\citenamefont{Ristroph, Bergou,
  Ristroph, Coumes, Berman, Guckenheimer, Wang, and Cohen}}]{RistrophPNAS2010}
\bibinfo{author}{\bibfnamefont{L.}~\bibnamefont{Ristroph}},
  \bibinfo{author}{\bibfnamefont{A.~J.} \bibnamefont{Bergou}},
  \bibinfo{author}{\bibfnamefont{G.}~\bibnamefont{Ristroph}},
  \bibinfo{author}{\bibfnamefont{K.}~\bibnamefont{Coumes}},
  \bibinfo{author}{\bibfnamefont{G.~J.} \bibnamefont{Berman}},
  \bibinfo{author}{\bibfnamefont{J.}~\bibnamefont{Guckenheimer}},
  \bibinfo{author}{\bibfnamefont{Z.~J.} \bibnamefont{Wang}}, \bibnamefont{and}
  \bibinfo{author}{\bibfnamefont{I.}~\bibnamefont{Cohen}},
  \bibinfo{journal}{Proceedings of the National Academy of Sciences}
  \textbf{\bibinfo{volume}{107}}, \bibinfo{pages}{4820} (\bibinfo{year}{2010}).

\bibitem[{\citenamefont{Pringle}(1948)}]{PringleGyroscopic1948}
\bibinfo{author}{\bibfnamefont{J.~W.~S.} \bibnamefont{Pringle}},
  \bibinfo{journal}{Philosophical Transactions of the Royal Society B:
  Biological Sciences} \textbf{\bibinfo{volume}{233}}, \bibinfo{pages}{347}
  (\bibinfo{year}{1948}).

\bibitem[{\citenamefont{Taylor}(2001)}]{TaylorMechanics2001}
\bibinfo{author}{\bibfnamefont{G.~K.} \bibnamefont{Taylor}},
  \bibinfo{journal}{Biological Reviews} \textbf{\bibinfo{volume}{76}},
  \bibinfo{pages}{449} (\bibinfo{year}{2001}).

\bibitem[{\citenamefont{Lourakis and Argyros}(2009)}]{LourakisSBA2009}
\bibinfo{author}{\bibfnamefont{M.~A.} \bibnamefont{Lourakis}} \bibnamefont{and}
  \bibinfo{author}{\bibfnamefont{A.}~\bibnamefont{Argyros}},
  \bibinfo{journal}{ACM Trans. Math. Software} \textbf{\bibinfo{volume}{36}},
  \bibinfo{pages}{1} (\bibinfo{year}{2009}).

\bibitem[{\citenamefont{Theriault et~al.}(2014)\citenamefont{Theriault, Fuller,
  Jackson, Bluhm, Evangelista, Wu, Betke, and Hedrick}}]{TheriaultJEB2014}
\bibinfo{author}{\bibfnamefont{D.~H.} \bibnamefont{Theriault}},
  \bibinfo{author}{\bibfnamefont{N.~W.} \bibnamefont{Fuller}},
  \bibinfo{author}{\bibfnamefont{B.~E.} \bibnamefont{Jackson}},
  \bibinfo{author}{\bibfnamefont{E.}~\bibnamefont{Bluhm}},
  \bibinfo{author}{\bibfnamefont{D.}~\bibnamefont{Evangelista}},
  \bibinfo{author}{\bibfnamefont{Z.}~\bibnamefont{Wu}},
  \bibinfo{author}{\bibfnamefont{M.}~\bibnamefont{Betke}}, \bibnamefont{and}
  \bibinfo{author}{\bibfnamefont{T.~L.} \bibnamefont{Hedrick}},
  \bibinfo{journal}{The Journal of Experimental Biology}
  \textbf{\bibinfo{volume}{217}}, \bibinfo{pages}{1843} (\bibinfo{year}{2014}).

\bibitem[{\citenamefont{Dickinson et~al.}(1999)\citenamefont{Dickinson,
  Lehmann, and Sane}}]{DickinsonScience1999}
\bibinfo{author}{\bibfnamefont{M.~H.} \bibnamefont{Dickinson}},
  \bibinfo{author}{\bibfnamefont{F.-O.} \bibnamefont{Lehmann}},
  \bibnamefont{and} \bibinfo{author}{\bibfnamefont{S.~P.} \bibnamefont{Sane}},
  \bibinfo{journal}{Science} \textbf{\bibinfo{volume}{284}},
  \bibinfo{pages}{1954} (\bibinfo{year}{1999}).

\bibitem[{\citenamefont{Sane and Dickinson}(2001)}]{SaneJEB2001}
\bibinfo{author}{\bibfnamefont{S.~P.} \bibnamefont{Sane}} \bibnamefont{and}
  \bibinfo{author}{\bibfnamefont{M.~H.} \bibnamefont{Dickinson}},
  \bibinfo{journal}{Journal of Experimental Biology}
  \textbf{\bibinfo{volume}{204}}, \bibinfo{pages}{2607} (\bibinfo{year}{2001}).

\bibitem[{\citenamefont{Ristroph et~al.}(2011)\citenamefont{Ristroph, Bergou,
  Guckenheimer, Wang, and Cohen}}]{RistrophPaddling2011}
\bibinfo{author}{\bibfnamefont{L.}~\bibnamefont{Ristroph}},
  \bibinfo{author}{\bibfnamefont{A.~J.} \bibnamefont{Bergou}},
  \bibinfo{author}{\bibfnamefont{J.}~\bibnamefont{Guckenheimer}},
  \bibinfo{author}{\bibfnamefont{Z.~J.} \bibnamefont{Wang}}, \bibnamefont{and}
  \bibinfo{author}{\bibfnamefont{I.}~\bibnamefont{Cohen}},
  \bibinfo{journal}{Phys. Rev. Lett.} \textbf{\bibinfo{volume}{106}},
  \bibinfo{pages}{178103} (\bibinfo{year}{2011}).

\bibitem[{\citenamefont{Sefati et~al.}(2013)\citenamefont{Sefati, Neveln, Roth,
  Mitchell, Snyder, MacIver, Fortune, and Cowan}}]{sefati2013mutually}
\bibinfo{author}{\bibfnamefont{S.}~\bibnamefont{Sefati}},
  \bibinfo{author}{\bibfnamefont{I.~D.} \bibnamefont{Neveln}},
  \bibinfo{author}{\bibfnamefont{E.}~\bibnamefont{Roth}},
  \bibinfo{author}{\bibfnamefont{T.~R.} \bibnamefont{Mitchell}},
  \bibinfo{author}{\bibfnamefont{J.~B.} \bibnamefont{Snyder}},
  \bibinfo{author}{\bibfnamefont{M.~A.} \bibnamefont{MacIver}},
  \bibinfo{author}{\bibfnamefont{E.~S.} \bibnamefont{Fortune}},
  \bibnamefont{and} \bibinfo{author}{\bibfnamefont{N.~J.} \bibnamefont{Cowan}},
  \bibinfo{journal}{Proceedings of the National Academy of Sciences}
  \textbf{\bibinfo{volume}{110}}, \bibinfo{pages}{18798}
  (\bibinfo{year}{2013}).

\bibitem[{\citenamefont{Cheng et~al.}(2009)\citenamefont{Cheng, Fry, Huang,
  Dickson, Dickinson, and Deng}}]{ChengIEEE2009}
\bibinfo{author}{\bibfnamefont{B.}~\bibnamefont{Cheng}},
  \bibinfo{author}{\bibfnamefont{S.}~\bibnamefont{Fry}},
  \bibinfo{author}{\bibfnamefont{Q.}~\bibnamefont{Huang}},
  \bibinfo{author}{\bibfnamefont{W.}~\bibnamefont{Dickson}},
  \bibinfo{author}{\bibfnamefont{M.}~\bibnamefont{Dickinson}},
  \bibnamefont{and} \bibinfo{author}{\bibfnamefont{X.}~\bibnamefont{Deng}}, in
  \emph{\bibinfo{booktitle}{Robotics and Automation, 2009. ICRA '09. IEEE
  International Conference on}} (\bibinfo{year}{2009}), pp.
  \bibinfo{pages}{1889--1896}, ISSN \bibinfo{issn}{1050-4729}.

\bibitem[{\citenamefont{Cheng and Deng}(2010)}]{Cheng2010Near}
\bibinfo{author}{\bibfnamefont{B.}~\bibnamefont{Cheng}} \bibnamefont{and}
  \bibinfo{author}{\bibfnamefont{X.}~\bibnamefont{Deng}}, pp.
  \bibinfo{pages}{39--44} (\bibinfo{year}{2010}).

\bibitem[{\citenamefont{Huston and Krapp}(2009)}]{Huston2009Gated}
\bibinfo{author}{\bibfnamefont{S.~J.} \bibnamefont{Huston}} \bibnamefont{and}
  \bibinfo{author}{\bibfnamefont{H.~G.} \bibnamefont{Krapp}},
  \bibinfo{journal}{The Journal of Neuroscience} \textbf{\bibinfo{volume}{29}},
  \bibinfo{pages}{13097} (\bibinfo{year}{2009}).

\bibitem[{\citenamefont{Zanker}(1988)}]{ZankerLongitudinal1988}
\bibinfo{author}{\bibfnamefont{J.~M.} \bibnamefont{Zanker}},
  \bibinfo{journal}{Physiological Entomology} \textbf{\bibinfo{volume}{13}},
  \bibinfo{pages}{351} (\bibinfo{year}{1988}), ISSN \bibinfo{issn}{1365-3032}.

\bibitem[{\citenamefont{Pringle}(1949)}]{PringleExcitation1949}
\bibinfo{author}{\bibfnamefont{J.}~\bibnamefont{Pringle}},
  \bibinfo{journal}{The Journal of physiology} \textbf{\bibinfo{volume}{108}},
  \bibinfo{pages}{226} (\bibinfo{year}{1949}).

\bibitem[{\citenamefont{Fox and Daniel}(2008)}]{FoxHaltere2008}
\bibinfo{author}{\bibfnamefont{J.}~\bibnamefont{Fox}} \bibnamefont{and}
  \bibinfo{author}{\bibfnamefont{T.}~\bibnamefont{Daniel}},
  \bibinfo{journal}{Journal of Comparative Physiology A}
  \textbf{\bibinfo{volume}{194}}, \bibinfo{pages}{887} (\bibinfo{year}{2008}),
  ISSN \bibinfo{issn}{0340-7594}.

\bibitem[{\citenamefont{Fayyazuddin and
  Dickinson}(1999)}]{FayyazuddinMuscle1999}
\bibinfo{author}{\bibfnamefont{A.}~\bibnamefont{Fayyazuddin}} \bibnamefont{and}
  \bibinfo{author}{\bibfnamefont{M.~H.} \bibnamefont{Dickinson}},
  \bibinfo{journal}{Journal of Neurophysiology} \textbf{\bibinfo{volume}{82}},
  \bibinfo{pages}{1916} (\bibinfo{year}{1999}), ISSN \bibinfo{issn}{0022-3077}.

\bibitem[{\citenamefont{Heide}(1983)}]{HeideNeural1983}
\bibinfo{author}{\bibfnamefont{G.}~\bibnamefont{Heide}},
  \bibinfo{journal}{BIONA-report} \textbf{\bibinfo{volume}{2}},
  \bibinfo{pages}{35} (\bibinfo{year}{1983}).

\bibitem[{\citenamefont{Heide and G{\"o}tz}(1996)}]{HeideOptomotor1996}
\bibinfo{author}{\bibfnamefont{G.}~\bibnamefont{Heide}} \bibnamefont{and}
  \bibinfo{author}{\bibfnamefont{K.~G.} \bibnamefont{G{\"o}tz}},
  \bibinfo{journal}{The Journal of experimental biology}
  \textbf{\bibinfo{volume}{199}}, \bibinfo{pages}{1711} (\bibinfo{year}{1996}).

\bibitem[{\citenamefont{Miyan and Ewing}(1985)}]{MiyanDiptera1985}
\bibinfo{author}{\bibfnamefont{J.~A.} \bibnamefont{Miyan}} \bibnamefont{and}
  \bibinfo{author}{\bibfnamefont{A.~W.} \bibnamefont{Ewing}},
  \bibinfo{journal}{Philosophical Transactions of the Royal Society B:
  Biological Sciences} \textbf{\bibinfo{volume}{311}}, \bibinfo{pages}{271}
  (\bibinfo{year}{1985}).

\bibitem[{\citenamefont{Dickinson and Tu}(1997)}]{TuDickinsonMuscle1997}
\bibinfo{author}{\bibfnamefont{M.~H.} \bibnamefont{Dickinson}}
  \bibnamefont{and} \bibinfo{author}{\bibfnamefont{M.~S.} \bibnamefont{Tu}},
  \bibinfo{journal}{Comparative Biochemistry and Physiology Part A: Physiology}
  \textbf{\bibinfo{volume}{116}}, \bibinfo{pages}{223} (\bibinfo{year}{1997}).

\bibitem[{\citenamefont{Walker et~al.}(2014)\citenamefont{Walker, Schwyn,
  Mokso, Wicklein, Müller, Doube, Stampanoni, Krapp, and
  Taylor}}]{WalkerMicrotomography2014}
\bibinfo{author}{\bibfnamefont{S.~M.} \bibnamefont{Walker}},
  \bibinfo{author}{\bibfnamefont{D.~A.} \bibnamefont{Schwyn}},
  \bibinfo{author}{\bibfnamefont{R.}~\bibnamefont{Mokso}},
  \bibinfo{author}{\bibfnamefont{M.}~\bibnamefont{Wicklein}},
  \bibinfo{author}{\bibfnamefont{T.}~\bibnamefont{Müller}},
  \bibinfo{author}{\bibfnamefont{M.}~\bibnamefont{Doube}},
  \bibinfo{author}{\bibfnamefont{M.}~\bibnamefont{Stampanoni}},
  \bibinfo{author}{\bibfnamefont{H.~G.} \bibnamefont{Krapp}}, \bibnamefont{and}
  \bibinfo{author}{\bibfnamefont{G.~K.} \bibnamefont{Taylor}},
  \bibinfo{journal}{PLoS Biol} \textbf{\bibinfo{volume}{12}},
  \bibinfo{pages}{e1001823} (\bibinfo{year}{2014}).

\bibitem[{\citenamefont{Balint and Dickinson}(2004)}]{BalintDickinsonJEB2004}
\bibinfo{author}{\bibfnamefont{C.~N.} \bibnamefont{Balint}} \bibnamefont{and}
  \bibinfo{author}{\bibfnamefont{M.~H.} \bibnamefont{Dickinson}},
  \bibinfo{journal}{Journal of experimental biology}
  \textbf{\bibinfo{volume}{207}}, \bibinfo{pages}{3813} (\bibinfo{year}{2004}).

\bibitem[{\citenamefont{Lin et~al.}(2014)\citenamefont{Lin, Ros, and
  Biewener}}]{lin2014through}
\bibinfo{author}{\bibfnamefont{H.-T.} \bibnamefont{Lin}},
  \bibinfo{author}{\bibfnamefont{I.~G.} \bibnamefont{Ros}}, \bibnamefont{and}
  \bibinfo{author}{\bibfnamefont{A.~A.} \bibnamefont{Biewener}},
  \bibinfo{journal}{Journal of The Royal Society Interface}
  \textbf{\bibinfo{volume}{11}}, \bibinfo{pages}{20140239}
  (\bibinfo{year}{2014}).

\bibitem[{\citenamefont{Hafting et~al.}(2005)\citenamefont{Hafting, Fyhn,
  Molden, Moser, and Moser}}]{Hafting2005Microstructure}
\bibinfo{author}{\bibfnamefont{T.}~\bibnamefont{Hafting}},
  \bibinfo{author}{\bibfnamefont{M.}~\bibnamefont{Fyhn}},
  \bibinfo{author}{\bibfnamefont{S.}~\bibnamefont{Molden}},
  \bibinfo{author}{\bibfnamefont{M.-B.} \bibnamefont{Moser}}, \bibnamefont{and}
  \bibinfo{author}{\bibfnamefont{E.~I.} \bibnamefont{Moser}},
  \bibinfo{journal}{Nature} \textbf{\bibinfo{volume}{436}},
  \bibinfo{pages}{801} (\bibinfo{year}{2005}).

\bibitem[{\citenamefont{Yartsev et~al.}(2011)\citenamefont{Yartsev, Witter, and
  Ulanovsky}}]{Yartsev2011Grid}
\bibinfo{author}{\bibfnamefont{M.~M.} \bibnamefont{Yartsev}},
  \bibinfo{author}{\bibfnamefont{M.~P.} \bibnamefont{Witter}},
  \bibnamefont{and}
  \bibinfo{author}{\bibfnamefont{N.}~\bibnamefont{Ulanovsky}},
  \bibinfo{journal}{Nature} \textbf{\bibinfo{volume}{479}},
  \bibinfo{pages}{103} (\bibinfo{year}{2011}).

\bibitem[{\citenamefont{Ofstad et~al.}(2011)\citenamefont{Ofstad, Zuker, and
  Reiser}}]{Ofstad2011Visual}
\bibinfo{author}{\bibfnamefont{T.~A.} \bibnamefont{Ofstad}},
  \bibinfo{author}{\bibfnamefont{C.~S.} \bibnamefont{Zuker}}, \bibnamefont{and}
  \bibinfo{author}{\bibfnamefont{M.~B.} \bibnamefont{Reiser}},
  \bibinfo{journal}{Nature} \textbf{\bibinfo{volume}{474}},
  \bibinfo{pages}{204} (\bibinfo{year}{2011}).

\bibitem[{\citenamefont{Seelig and Jayaraman}(2013)}]{Seelig2013Feature}
\bibinfo{author}{\bibfnamefont{J.~D.} \bibnamefont{Seelig}} \bibnamefont{and}
  \bibinfo{author}{\bibfnamefont{V.}~\bibnamefont{Jayaraman}},
  \bibinfo{journal}{Nature} \textbf{\bibinfo{volume}{503}},
  \bibinfo{pages}{262} (\bibinfo{year}{2013}).

\bibitem[{\citenamefont{Fox et~al.}(2010)\citenamefont{Fox, Fairhall, and
  Daniel}}]{Fox2010Encoding}
\bibinfo{author}{\bibfnamefont{J.~L.} \bibnamefont{Fox}},
  \bibinfo{author}{\bibfnamefont{A.~L.} \bibnamefont{Fairhall}},
  \bibnamefont{and} \bibinfo{author}{\bibfnamefont{T.~L.}
  \bibnamefont{Daniel}}, \bibinfo{journal}{Proceedings of the National Academy
  of Sciences} \textbf{\bibinfo{volume}{107}}, \bibinfo{pages}{3840}
  (\bibinfo{year}{2010}).

\end{thebibliography}


\begin{thebibliography}{3}
\expandafter\ifx\csname natexlab\endcsname\relax\def\natexlab#1{#1}\fi
\expandafter\ifx\csname bibnamefont\endcsname\relax
  \def\bibnamefont#1{#1}\fi
\expandafter\ifx\csname bibfnamefont\endcsname\relax
  \def\bibfnamefont#1{#1}\fi
\expandafter\ifx\csname citenamefont\endcsname\relax
  \def\citenamefont#1{#1}\fi
\expandafter\ifx\csname url\endcsname\relax
  \def\url#1{\texttt{#1}}\fi
\expandafter\ifx\csname urlprefix\endcsname\relax\def\urlprefix{URL }\fi
\providecommand{\bibinfo}[2]{#2}
\providecommand{\eprint}[2][]{\url{#2}}

\bibitem[{\citenamefont{Chang and Wang}(2014)}]{ChangPNAS2014}
\bibinfo{author}{\bibfnamefont{S.}~\bibnamefont{Chang}} \bibnamefont{and}
  \bibinfo{author}{\bibfnamefont{Z.~J.} \bibnamefont{Wang}},
  \bibinfo{journal}{Proceedings of the National Academy of Sciences}
  \textbf{\bibinfo{volume}{111}}, \bibinfo{pages}{11246}
  (\bibinfo{year}{2014}).

\bibitem[{\citenamefont{Dickinson}(1999)}]{DickinsonHaltere1999}
\bibinfo{author}{\bibfnamefont{M.~H.} \bibnamefont{Dickinson}},
  \bibinfo{journal}{Philosophical Transactions of the Royal Society of
  London.Series B: Biological Sciences} \textbf{\bibinfo{volume}{354}},
  \bibinfo{pages}{903} (\bibinfo{year}{1999}).

\bibitem[{\citenamefont{Beatus et~al.}(2015)\citenamefont{Beatus, Guckenheimer,
  and Cohen}}]{BeatusInterface2015}
\bibinfo{author}{\bibfnamefont{T.}~\bibnamefont{Beatus}},
  \bibinfo{author}{\bibfnamefont{J.~M.} \bibnamefont{Guckenheimer}},
  \bibnamefont{and} \bibinfo{author}{\bibfnamefont{I.}~\bibnamefont{Cohen}},
  \bibinfo{journal}{Journal of The Royal Society Interface}
  \textbf{\bibinfo{volume}{12}} (\bibinfo{year}{2015}), ISSN
  \bibinfo{issn}{1742-5689}.

\end{thebibliography}

\end{document}


\title{Pitch Perfect: How Fruit Flies Control their Body Pitch Angle \\ Supplementary Information}
 
\author{Samuel C. Whitehead$^{1*}$, Tsevi Beatus$^{1*}$, Luca Canale$^2$, and Itai Cohen$^1$}
\affiliation{$^1$Department of Physics, Cornell University, Ithaca, New York 14853, USA; $^2$D\'{e}partement de M\'{e}canique, \'{E}cole Polytechnique 911128, Palaiseau, France. \\ $^*$ Equal contributors }

\date{\today}

\maketitle

\section{Movies}

\emph{Movie 1:-} A fruit fly undergoing a typical pitch up perturbation and correction maneuver, corresponding to the data presented in Figures 1-4 and 6 in the main text. The side panels of the 3D box correspond to raw movies from the three orthogonal high-speed cameras. The green line, and its white projection onto the floor of the 3D box, show the center-of-mass trajectory of the fly, with red regions of the line representing the duration of the magnetic pulse. The timing for the event, in milliseconds, is given in the bottom left corner. If not attached, the video can be found at \url{https://youtu.be/2FGj7HUCL7E}. 

\vspace{5mm}

\emph{Movie 2:-} Raw footage of three camera views of a fly undergoing a large-amplitude pitch down perturbation, corresponding to the data presented in Figure 8. The center panel corresponds to the overhead view, while the left and right panels correspond to side views. The timing for the event, in milliseconds, is given in the bottom left corner. The magnetic perturbation is applied from 0-5.8 ms. If not attached, the video can be found at \url{https://youtu.be/EUIE9jRcusg}.

\section{Simplified Wing Kinematics}

For both the calculation of aerodynamic torques in Figure 5a and the dynamical simulation in Figure 7, we use an analytic form for simplified wing kinematics taken from \cite{ChangPNAS2014}. These kinematics closely resemble the motion of real fly wings, but are simple enough to write down concisely. 
\begin{align}
	\phi_\text{w}(t) &= \phi_\text{0} + \phi_\text{m}\frac{\text{asin}(K\text{sin}(\omega t))}{\text{asin}(K)} \label{eq:phi} \\
	\eta_\text{w}(t) &= \eta_\text{0} + \eta_\text{m}\frac{\text{tanh}(C\text{sin}(\omega t + \delta_\eta))}{\text{tanh}(C)} \label{eq:eta}  \\
	\theta_\text{w}(t) &= \theta_\text{0} + \theta_\text{m}\text{cos}(2\omega t + \delta_\theta) \label{eq:theta}
\end{align}
The wing Euler angles are defined in Figure 1c in the main text. The terms in Equations \ref{eq:phi}, \ref{eq:eta}, and \ref{eq:theta} are defined as follows:
\begin{itemize}
	\item $\phi_\text{0}$, $\eta_\text{0}$, $\theta_\text{0}$ are angle offsets.
	\item $\phi_\text{m}$, $\eta_\text{m}$, $\theta_\text{m}$ are amplitudes.
	\item $K$, $C$ are waveform parameters. $K$ tunes the stroke angle from pure sine wave to triangle wave, while $C$ tunes the wing pitch angle from sinusoid to square wave.
	\item $\omega$ is the wing-beat frequency. 
	\item $\delta_\eta$, $\delta_\theta$ are phase offsets.
\end{itemize}
For the calculation in Figure 5a we use $\omega = 250\times(2\pi)$ rad/s; $\eta_\text{m} = 53\degree$; $\eta_\text{0} = 90\degree$ ; $\delta_\eta = 72.4\degree$ ; $C$ = 2.4; $\theta_\text{m} = 25\degree$; $\theta_\text{0} = 0\degree$; $\delta_\theta = 90\degree$; $K = .7$; $\phi_\text{m} \in [60\degree, 87.5\degree]$; $\phi_\text{0} \in [87.5\degree, 115\degree]$. For the simulation in Figure 6, we use similar parameters to the above, but set $\theta_\text{m} = \theta_\text{0} = 0\degree$ for simplicity and vary $\phi_\text{m}$ and $\phi_\text{0}$ according to our controller model. Note that in the main text we refer to front and back stroke angles ($\phi_\text{w}^\text{front}$ and $\phi_\text{w}^\text{back}$), which are related to $\phi_\text{m}$ and $\phi_\text{0}$ by the linear relations:
\begin{align}
	\phi_\text{w}^\text{front} &= \phi_\text{0} - \phi_\text{m} \\
	\phi_\text{w}^\text{back} &= \phi_\text{0} + \phi_\text{m}
\end{align}

\section{Numerical Simulation}

Using the simplified wing kinematics above and the quasi-steady aerodynamic model detailed in the main text (Equations 1 and 2), our numerical simulation solves the Newton-Euler equations for vertical, forward, and pitch rotational motion. Control is implemented by adjusting the front stroke angle of the prescribed wing kinematics according to Equation 3 in the main text. The $K_\text{i}$ and $K_\text{p}$ parameters were determined by the fit to the experimental data. The inputs for the controller in each wing beat--the body pitch angle and pitch velocity--were taken as their mean values during the previous wing beat. This scheme represents a time-delay of one wing-beat while avoiding the effects of the inherent small-scale pitch oscillations. Before we apply the perturbation, we let the simulated fly stabilize to a steady-state body pitch angle of 45$\degree$. The perturbation is then applied, with magnitude roughly corresponding to the accelerations observed in the experiments. In simulation runs that tested controller models other than PI, we let the system stabilize at $\theta_\text{b}$ = 45$\degree$ using a PI controller and only then we applied the perturbation and simultaneously changed the controller type.

\begin{figure*}[t]
	\centering
		\includegraphics[scale = .9]{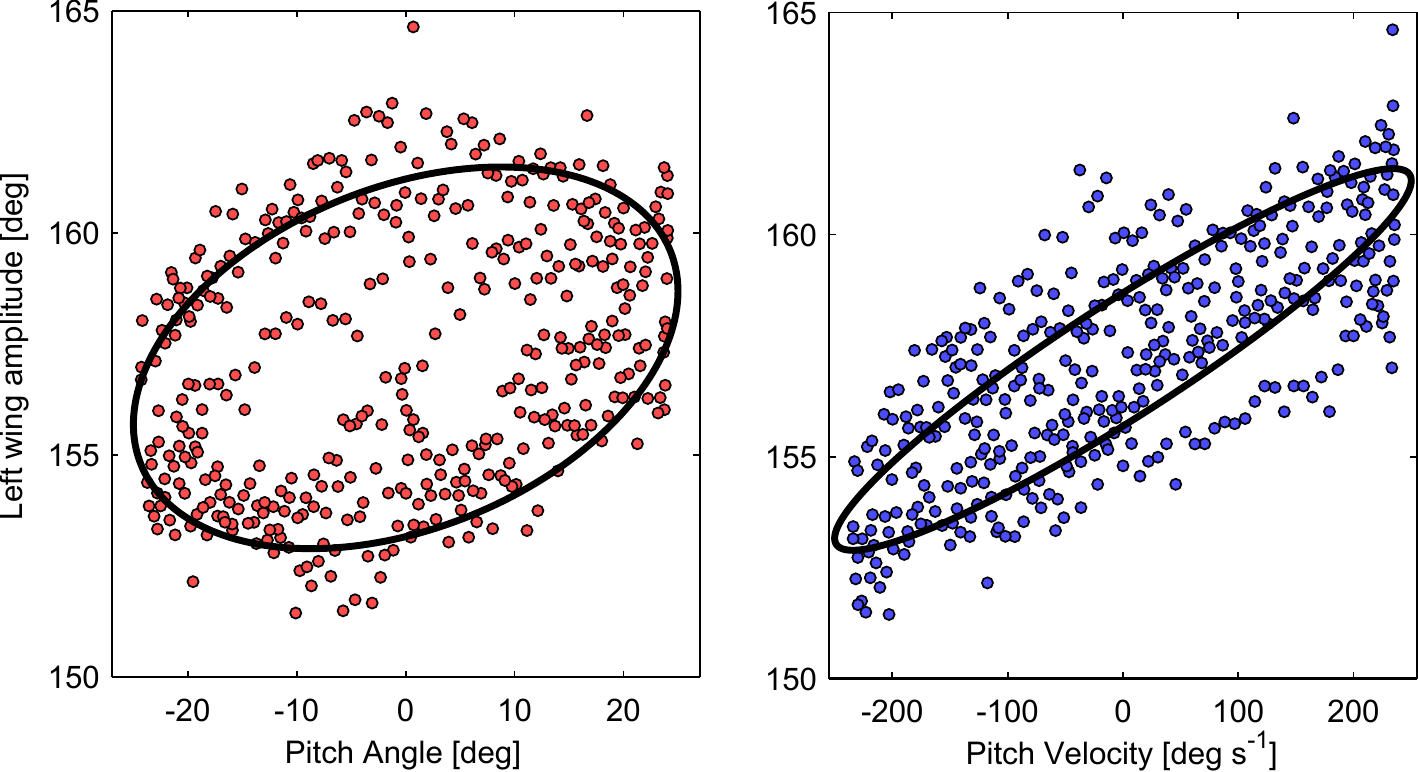}
	  \caption{Data extracted from \cite{DickinsonHaltere1999} for the left wing amplitude in response to sinusoidal pitch perturbation. On the top,
the red circles represent the response to the pitch angle; on the bottom panel, the blue circles represent the response to the pitch
velocity in the same measurement. The black solid ellipses show the predicted response of a PI control model.}
	 \label{fig:dickinson1}
\end{figure*}

\section{Analysis of Data from Previous Studies}

Following \cite{BeatusInterface2015}, we further test the PI control model by using it to predict the pitch response of tethered fruit flies previously published by Dickinson \cite{DickinsonHaltere1999}. In \cite{DickinsonHaltere1999}, flies were tethered to a gimbal apparatus that oscillated about different rotational axes. The left wing stroke amplitude of the flies, $\Phi_\text{left}$, was measured using photodetectors that recorded the wings' shadows. As noted by \cite{DickinsonHaltere1999}, flies do not adjust their back stroke angle during pitch correction, so stroke amplitude is a good proxy for $\phi_\text{w}^\text{front}$, the quantity we measure in free flight experiments. 

In one of the measurements reported in \cite{DickinsonHaltere1999}, pitch perturbations were imposed so that: $\theta_\text{b}(t) = A\text{sin}(\omega t)$ and $\dot{\theta}_\text{b}(t) = A\omega\text{cos}(\omega t)$, with $A = 25\degree$, period $T = 0.63$ s, and maximum pitch velocity $250\degree$s$^{-1}$. The left wing stroke amplitude was plotted against both pitch angle and pitch velocity (Figure 3a,c in \cite{DickinsonHaltere1999}). Using standard image processing techniques, we extract the data from these plots.

The pitch oscillations in the tethered experiments have period 630 ms, which is much longer than the observed pitch correction latency times from our experiments ($\approx 10$ ms), so we consider the controller time delay negligible. We then write the form for our controller, now in terms of left wing-stroke amplitude, as:
\begin{equation}
\label{eq:ctrl2}
\Phi_\text{left}(t) =  K_\text{p} \dot{\theta}_\text{b}(t) + K_\text{i}\Delta\theta_\text{b}(t) + \Phi_\text{mean}
\end{equation} 
where the left wing stroke amplitude, $\Phi_\text{left}(t)$, and mean stroke amplitude, $\Phi_\text{mean}$, are related to the controlled quantity from the main text, $\Delta\phi_\text{w}^\text{front}(t)$, by the linear relation: $\Delta\phi_\text{w}^\text{front}(t) = \Phi_\text{left}(t) - \Phi_\text{mean}$. Note that considering only the left wing does not reduce generality of this analysis, since pitch correction is left/right symmetric. We manually fit for for the control parameters from the data. The fitted parameters obtained are $K_\text{i} = 0.3$ and $K_\text{p} = 8$ ms, comparable to the parameters from the main text.

\begin{figure}
	\centering
		\includegraphics[scale = .95]{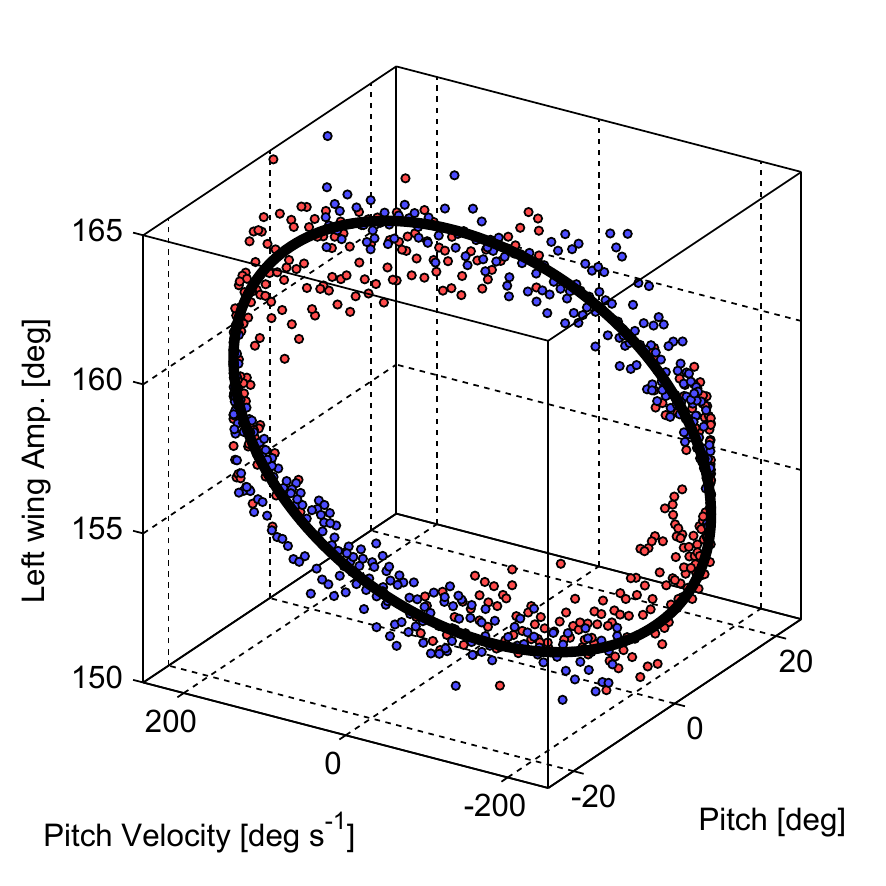}
	  \caption{Data extracted from \cite{DickinsonHaltere1999} for the left wing amplitude in response to sinusoidal pitch perturbation is plotted in the 3D space whose axes are ($\theta_\text{b}$, $\dot{\theta}_\text{b}$, $\Phi_\text{left}$). The data points are colored according to their original color in Figure \ref{fig:dickinson1}. The black solid line shows the predicted response of the same PI
control model shown in Figure \ref{fig:dickinson1}.}
	 \label{fig:dickinson2}
\end{figure}

\begin{figure}
	\centering
		\includegraphics[scale = .95]{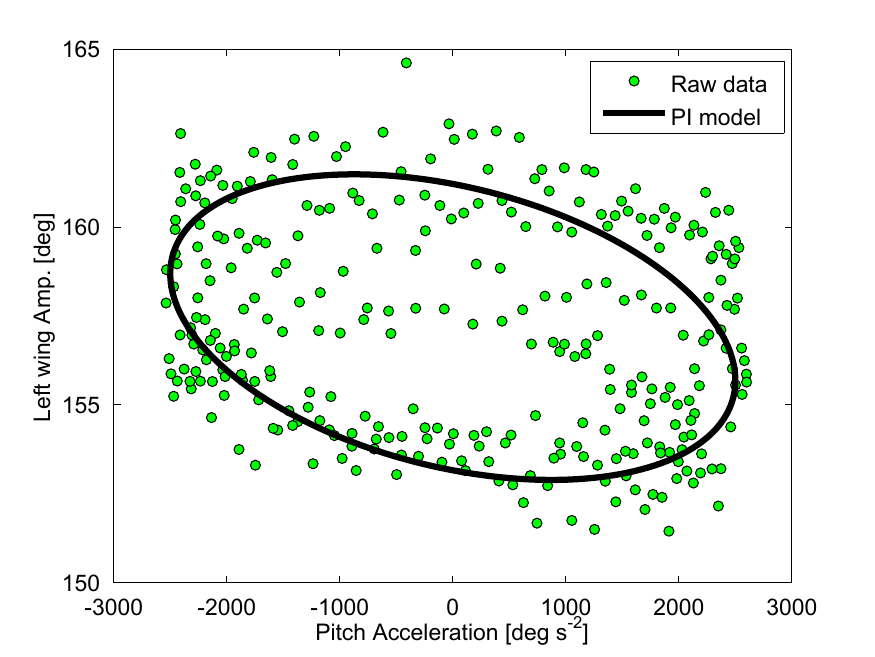}
	  \caption{Data extracted from \cite{DickinsonHaltere1999} for the left wing amplitude in response to sinusoidal pitch perturbation is plotted as a function of pitch acceleration. The data corresponds to the same measurement (same fly and same perturbation) as shown in Figures \ref{fig:dickinson1} and \ref{fig:dickinson2}. The black solid line shows the predicted response of the same PI control model as in Figures \ref{fig:dickinson1} and \ref{fig:dickinson2}. Note that the PI controller model includes no information about the pitch acceleration, suggesting that pitch acceleration is unimportant in determining the flies' corrective wing kinematics, and thus excluding a PID controller model.}
	 \label{fig:dickinson3}
\end{figure}

The predictions of the PI controller fit are shown in Figures \ref{fig:dickinson1}, \ref{fig:dickinson2}, and \ref{fig:dickinson3}. The output of the PI controller is plotted as a function of both pitch angle and pitch velocity in Figure \ref{fig:dickinson1}, yielding an ellipse in both cases (R$^2$ = 0.761 and 0.842 respectively). The linear model for $\Phi_\text{left}$ as a function of $\dot{\theta}_\text{b}$ gives R$^2$ =  0.556. To show the full dependence of the PI controller on both angle and velocity, we also plotted the PI controller prediction in the 3D space whose axes are ($\theta_\text{b}$, $\dot{\theta}_\text{b}$, $\Phi_\text{left}$), shown in Figure \ref{fig:dickinson2}. The PI controller predicts an inclined ellipse in this space, the projections of which onto the horizontal axes yield the plots in Figure \ref{fig:dickinson1}. The inclination of the ellipse shows that corrective response depends on both pitch angle and pitch velocity, i.e. the inclination of the measured data in \cite{DickinsonHaltere1999} is consistent with a PI controller.

Additionally, we show the predicted output of the PI controller plotted as a function of pitch acceleration in Figure \ref{fig:dickinson3}. Consistent with \cite{DickinsonHaltere1999}, Figure \ref{fig:dickinson3} shows that the fly's corrective response can be quantitatively captured without including information about the pitch acceleration. Figure \ref{fig:dickinson3} suggests that pitch acceleration is unimportant in determining the flies' corrective wing kinematics, and thus excludes a PID controller model.

\bibliography{FlySources}